**FRONT MATTER**

# Feasibility of keeping Mars warm with nanoparticles

Short title: Feasibility of warming Mars with nanoparticles

**Authors**
Samaneh Ansari[1], Edwin S. Kite[2,*], Ramses Ramirez[3], Liam J. Steele[2,4], Hooman Mohseni[1]

**Affiliations**
1. Department of Electrical and Computer Engineering, Northwestern University; Evanston, IL.
2. Department of the Geophysical Sciences, University of Chicago; Chicago, IL.
3. Department of Physics, University of Central Florida; Orlando, FL.
4. European Center for Medium-Range Weather Forecasts; Reading, UK.
* Corresponding author, kite@uchicago.edu

**Abstract**
One-third of Mars' surface has shallow-buried $H_2O$, but it is currently too cold for use by life. Proposals to warm Mars using greenhouse gases require a large mass of ingredients that are rare on Mars' surface. However, we show here that artificial aerosols made from materials that are readily available at Mars—for example, conductive nanorods that are ~9 µm long—could warm Mars >5×10$^3$ times more effectively than the best gases. Such nanoparticles forward-scatter sunlight and efficiently block upwelling thermal infrared. Similar to the natural dust of Mars, they are swept high into Mars' atmosphere, allowing delivery from the near-surface. For a particle lifetime of 10 years, two climate models indicate that sustained release at 30 liters/sec would globally warm Mars by ≳30 K and start to melt the ice. Therefore, if nanoparticles can be made at scale on (or delivered to) Mars, then the barrier to warming of Mars appears to not be as high as previously thought.

**Teaser**
Warming Mars with artificial aerosols appears to be feasible.

**MAIN TEXT**

## Introduction.

Dry river valleys cross Mars's once-habitable surface (1, 2), but today the icy soil is too cold for Earth-derived life (3-5). Streams may have flowed as recently as 600 kyr ago (6), hinting at a planet on the cusp of habitability. Many methods have been proposed to warm Mars' surface by closing the spectral windows, centered around wavelengths ($\lambda$) 22 µm and 10 µm, through which the surface is cooled by thermal infrared radiation upwelling to space (7-9). Modern Mars has a thin (~6 mbar) $CO_2$ atmosphere that provides only ~5 K greenhouse warming via absorption in the 15 µm band (10), and Mars apparently lacks enough condensed or mineralized $CO_2$ to restore a warm climate (11). The spectral windows can be closed using artificial greenhouse gases (e.g.

chloroflourocarbons) (8, 12), but this would require volatilizing ~100,000 megatons of fluorine, which is sparse on the Mars surface. An alternative approach is suggested by natural Mars dust aerosol. Mars dust is almost all ultimately sourced from slow comminution (indirect estimate $O(3)$ liters/second, ref. 13) of iron-rich minerals on Mars' surface. Due to its small size (1.5 µm effective radius), Mars dust is lofted to high altitude (altitude of peak dust mass mixing ratio 15-25 km), is always visible in the Mars sky, and is present up to >60 km altitude (14-15). Natural Mars dust aerosol lowers daytime surface temperature (e.g., 16), but this is due to compositional and geometric specifics that can be modified in the case of engineered dust. For example, a nanorod about half as long as the wavelength of upwelling thermal infrared radiation should interact strongly with that radiation (17).

**Results.**

Consider a 9-µm long conductive nanorod (we consider aluminum and iron) with a ~60:1 aspect ratio, not much smaller than commercially available glitter. Finite-difference time domain calculations (Supplementary Methods) show that such nanorods, randomly oriented due to Brownian motion (18), would strongly scatter and absorb upwelling thermal infrared in the spectral windows, and forward-scatter sunlight down to the surface, leading to net warming (Fig. 1, Figs. S1-S4). Results are robust to changing particle material type, cross-sectional shape, and mesh resolution, and change as expected with particle length and aspect ratio (Figs. S5-S8). The calculated thermal infrared scattering is near-isotropic (Fig. 1), which favors surface warming (19). Such nanorods would settle >10× more slowly in the Mars atmosphere than natural Mars dust (Supplementary Methods), implying that once the particles are lifted into the air they would be lofted to high altitude and have a long atmospheric lifetime.

This motivates calculating surface warming (K) as a function of (artificial) aerosol column density (kg/m$^2$). The MarsWRF global climate model is suitable for such a calculation (20-22). Following many previous works (22-26), we prescribe a layer of aerosol, and calculate the resulting steady-state warming (Supplementary Methods). Our calculation does not include dynamic transport of aerosol, but includes realistic topography, seasonal forcing, and surface thermophysical properties and albedo. The model output (Fig. 2, Figs. S9-S19) shows that Al nanorod column density 160 mg/m$^2$ yields surface temperatures and pressures permitting extensive summertime (i.e., the warmest ~70 sols period each year) liquid water in locations with shallow ground-ice. This is >5000× more effective, on a warming-per-unit-mass-in-the-atmosphere basis, than the current state of the art (8) (Supplementary Methods). Temperatures experienced by subsurface ice will be lower due to insulation by soil. Water ice at <1 m depth is almost ubiquitous poleward of ±50° latitude (blue lines in Fig. 2) (1). $H_2O$ ice is present further equatorward (27), but is insulated beneath >1 m soil cover and so would not melt unless the annual average surface temperature is raised close to 273K.

The greenhouse effect depends on the temperature difference between the top of the optically thick IR emitting/absorbing layer and that of the planet surface; higher clouds have a bigger $\Delta T$ relative to the surface (due to adiabatic cooling) and so give a stronger greenhouse effect. Therefore, the results depend on both artificial-dust-layer-top height and column density (28). Supplementary

Table 2 and Fig. S19 show the results varying the layer-top height between ~35 km and ~28 km. The minimum column density for substantial warming (Fig. 2c) can be estimated by setting optical depth in the spectral window ($\tau_{win}$) to unity and solving the following expression for column density $M_c$ (mg/m$^2$) (e.g. 24):

$$\tau_{win} = 3\, Q_{eff}\, M_c \,/\, (4\, r\, \rho) \qquad (1)$$

Here, $Q_{eff}$ is the wavenumber-dependent extinction efficiency, $\rho$ is the nanorod particle density (Al: 2.7 g/cc), and $r$ is the effective nanorod particle radius (the radius of a sphere of equivalent volume, 0.38 µm). Here, $Q_{eff}$ is the ratio of the extinction cross-section in the spectral window (about one-half the maximum cross-section, i.e. $3 \times 10^{-11}$ m$^2$, from Fig. 1) to the geometric cross-section of the equivalent sphere, and is ≈60. This gives a minimum column density ($M_c$) of 20 mg/m$^2$. At 160 mg/m$^2$, the volumetric density of nanorods, 10 cm$^{-3}$, gives a Brownian coagulation timescale (for 0.1-10 µm diameter spheres, for 100% accretion efficiency) ≈ 6 years (18). This timescale estimate has substantial uncertainties; for example, actual accretion efficiency may be less, for example because monodisperse particles of uniform composition (e.g. nanorods) may carry similar charges and thus repel each other (29). On Mars, particles would be taken up by dry deposition and by transient $CO_2$ ice, and re-released to the atmosphere by dust lifting. Initial release (after manufacture) could be from a pipe extending 10-100m above the surface, as Mars turbulent updrafts strengthen with distance from the surface (30). For an effective particle lifetime of 10 years, sustaining the warming shown in Fig. 2a requires particle fountaining at an average rate of 30 liters/sec (1 liter/sec corresponds to the flow of one standard garden sprinkler). Multi-year lifetimes are consistent with one-pass fall-out of ~0.1 µm-diameter particles (31), and particle lifetime might be greatly extended if the particles are engineered to self-loft (32-34), further reducing the sustaining mass flux; however, effective particle lifetime remains a major uncertainty in our model.

As a check on the 3-D results, we ran a 1-D model using annual-average Mars insolation (Supplementary Methods). This predicts 245 K global temperature for ~160 mg/m$^2$ Al nanorods (Fig. 2c, Figs. S16-S17). For further increases in nanorod loading, the 1-D model predicts lower global temperatures than does the 3-D model. This may be due to differences in the vertical temperature structure of the two models (Supplementary Information).

Even in the warmed climate, the south pole is cold enough for seasonal $CO_2$ condensation.

Within months of warming Mars, the atmospheric pressure increases by ~20% as $CO_2$ ice sublimes, a positive warming feedback. On a warmed Mars, atmospheric pressure will further increase by a factor of 2-20 as adsorbed $CO_2$ desorbs (35), and polar $CO_2$ ice (36) is volatilized on a timescale that could be as long as centuries. This will further increase the area that is suitable for liquid water (6). However, raising Mars' temperature, by itself, is not sufficient to make the planet's surface habitable for oxygenic photosynthetic life: barriers remain (7). For example, Mars' sands have ~300 ppmw nitrates (37), and Mars' air contains very little $O_2$, as did Earth's air prior to the arrival of cyanobacteria. Remediating perchlorate-rich soil might require bioremediation by perchlorate-reducing bacteria, which yield molecular oxygen as a natural byproduct (38).

**Discussion.**
The results from this relatively simple workflow are subject to several uncertainties that motivate more sophisticated modeling. As one of several examples, modeling of coupled dust flow and ice nucleation on Mars is currently at an early stage (39). Modeling the effect of nanorods as ice nuclei - which could either be a positive or a negative feedback, depending on the size and altitude of the resulting water ice cloud particles and their precipitation efficiency - is additional motivation to study this coupling. A thin coating on the nanorods could alter their hydrophobicity level, and potentially the ice nucleation, and might also protect against oxidation. The optimal location(s) for particle fountaining require further research. Release into the ascending limb of the Hadley cell should allow dispersal in both hemispheres. The radiative effect of water vapor feedback is unambiguously positive. Tests varying nanorod size, composition, and shape suggest that further improvements to warming effectiveness are possible (Figs. S7-S8). For example, extinction efficiency decreases approximately linearly with rod radius, but mass decreases quadratically with rod radius.

With the caveats above in mind, Fig. 2c allows a first estimate of how much surface material would be needed to supply the fountains. For surface-material density 2 g/cc, and $Al_2O_3$ content of ~10 wt% (e.g., 40), raising the surface temperature to that shown in Fig. 2a over 10 years would require processing $2 \times 10^7$ $m^3$/yr surface material to obtain $7 \times 10^5$ $m^3$/yr of metal, corresponding to a prismatic mine of half-width 350m and side-wall slope 20°, lengthening by 250m per year. This is much easier than the current state-of-the-art (8), because fluorine is only infrequently detected by rovers (41), and is present in most Mars meteorites only at 15-90 ppmw concentrations (42). However, even this reduced material-processing demand still corresponds to $1 \times 10^{-3}$ of Earth's metal production, and this defines a major manufacturing problem that remains to be solved. Processing of surface material into nanoparticles might use lenses or mirrors to concentrate sunlight for vacuum evaporation, followed by colloidal growth. Synthetic biology (e.g. magnetite nanorods) is a possible alternative (43). Metal 3D printing of parts (e.g. Relativity's Stargate; 44) and/or assembly on Mars might reduce launch costs. Due to their <2 nm width, carbon nanomaterials might warm Mars more effectively than the nanoparticles shown in Figs. 1-2. For example, graphene's density is 0.77 mg/$m^2$, and graphene nanodisks that are ~$10^2$ nm diameter have strong mid-infrared resonances (45).

Although nanoparticles could warm Mars (Fig. 2c, Fig. 3), both the benefits and potential costs of this course of action are currently uncertain. For example, in the unlikely event that Mars' soil contains irremediable compounds toxic to all Earth-derived life (this can be tested with Mars Sample Return), then the benefit of warming Mars is nil. On the other hand, if a photosynthetic biosphere can be established on the surface of Mars, perhaps with the aid of synthetic biology, that might increase the Solar System's capacity for human flourishing. On the costs side, if Mars has extant life, then study of that life could have great benefits that warrant robust protections for its habitat. More immediately, further research into nanoparticle design and manufacture coupled with modeling of their interaction with the climate could reduce the expense of this method. Examples include Mars-pressure wind-tunnel experiments for nanorod and dust re-uptake and release rate from realistic rough surfaces (including icy surfaces), and mesoscale/large-eddy-simulation modeling of nanorod dispersal and lofting. In addition to the nanorod warming option and the no-action option, cost-benefit calculations should also consider local-warming methods, such as silica aerogel tiling (9).

More work is needed on the very-long-term sustainability of a warmed Mars. Atmospheric escape to space would take at least 300 Myr to deplete the atmosphere at the present-day rate (46). However, if the ground ice observed at meters to tens of meters depth is underlain by empty pore space, then excessive warming over centuries could allow water to drain away, requiring careful management of long-term warming. Subsurface exploration by electromagnetic methods could address this uncertainty regarding how much water remains on Mars deep underground (47).

The efficiency of nanoparticle warming suggests that any entity with the goal of strong planet-scale warming would use this approach. This suggests polarization as a technosignature for cold terrestrial worlds with geodynamos.

Nanoparticle warming, by itself, is not sufficient to make the planet's surface habitable again. Nevertheless, our study suggests that nanoparticle warming may be of interest to the nanophotonics and planetary science communities, among others, and that further investigation might be fruitful.

**Materials and Methods.**

*Calculation of optical properties of nanorods.* We carried out finite-difference time domain simulations (FDTD: 3D Electromagnetic Simulator). Optical properties are obtained at 75 wavelengths, approximately log-uniformly spaced from 0.24-55 µm (Table S1). Additional details are given in the Supplementary Materials.

*Calculation of the surface-warming effect of the nanorods using 1D climate model.* Our single-column radiative-convective climate model (RCM) subdivides atmospheres into multiple vertical log-layers (we implement 201 layers here) that extend from the ground to the top of the atmosphere ($1 \times 10^{-5}$ bar here) (e.g. 25). The RCM has 55 infrared and 38 solar spectral intervals and applies a standard moist convective adjustment. Additional details are given in the Supplementary Materials.

*Calculation of the surface-warming effect of the nanorods using 3D climate model.* We use the MarsWRF code (20-22), with a horizontal resolution of 5.625° × 3.75°, corresponding to a grid of 64 points in longitude × 48 points in latitude. A 40-layer vertical grid is used. A prescribed natural dust aerosol distribution is imposed, giving an average dust optical depth of 0.20 (peaking at 0.43 during southern summer). Additional details are given in the Supplementary Materials.

**Competing interests:** Authors declare that they have no competing interests.

**Data and materials availability.** All data needed to evaluate the conclusions in the paper are present in the paper and/or the Supplementary Materials. Additional data, for example full 3D climate model output files, are stored at Zenodo (doi:10.5281/zenodo.8352416). FDTD:

3D Electromagnetic Simulator is commercial code (Lumerical). The MarsWRF source code can be made available by Aeolis Research pending scientific review and a completed Rules of the Road agreement. Requests for the MarsWRF source code should be submitted to mir@aeolisresearch.com.

**References.**

**Acknowledgements.**
We thank M. Turbet, C. Willard, M.A. Mischna, A. Geller, M.I. Richardson, C. Lee, F. Sharipov, A. Noblet, and the PlanetWRF development team. This research was supported in part through the computational resources and staff contributions provided for the Quest high performance computing facility at Northwestern University which is jointly supported by the Office of the Provost, the Office for Research, and Northwestern University Information Technology, and with resources provided by the University of Chicago's Research Computing Center.
*Funding:* The authors acknowledge that they received no external funding in support of this research.

**Author contributions.** ESK conceived research. SA, ESK, RR, LJS, and HM designed research. SA, ESK, and RR carried out research. ESK drafted the manuscript. SA, ESK, RR, LJS, and HM edited the manuscript and contributed to the final manuscript.

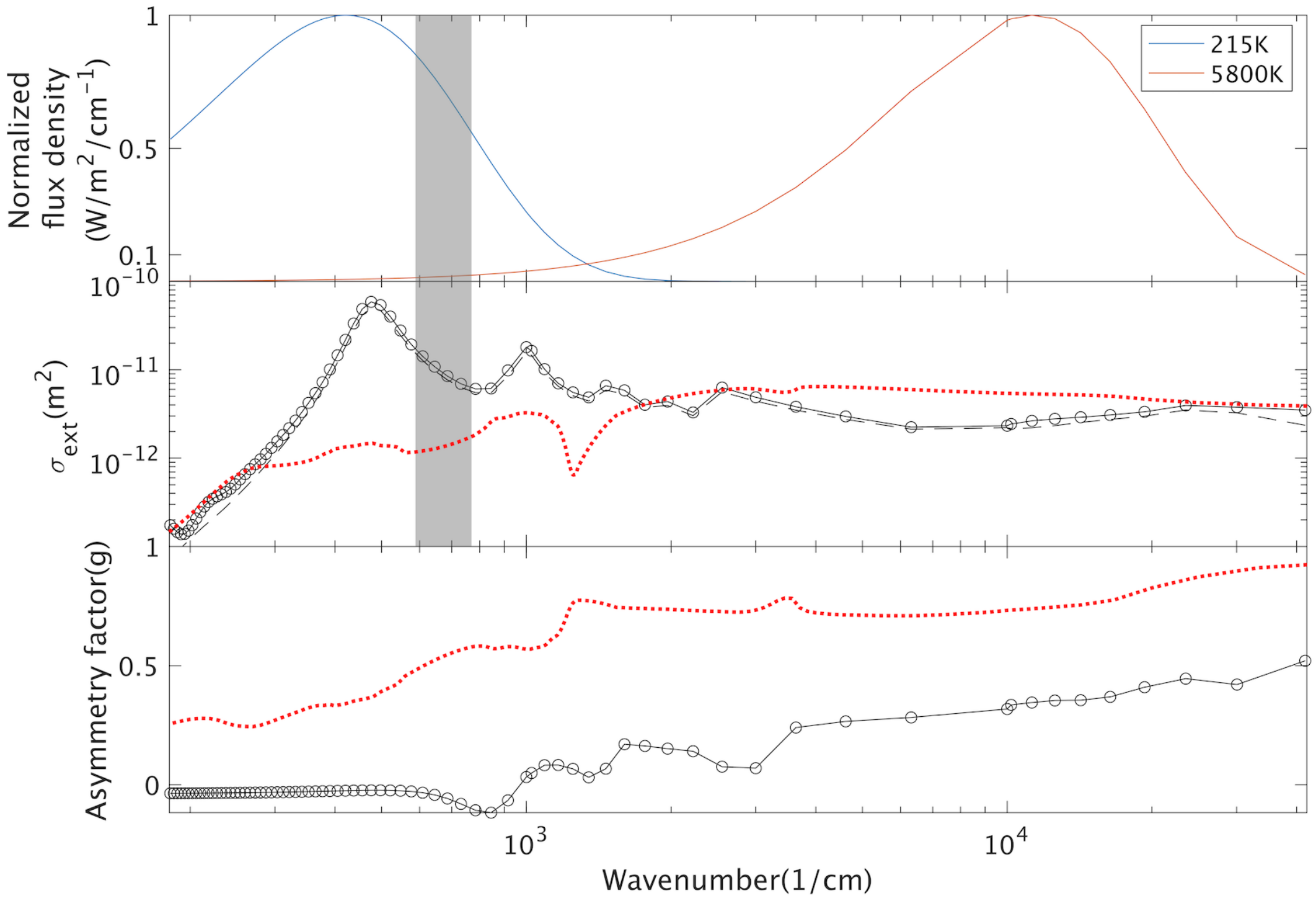

**Fig. 1. Orientation-averaged optical properties of a 9 µm-long Al nanorod with cross-section 0.16 µm × 0.16 µm, calculated using a 3D Finite-Difference Time-Domain (FDTD) approach.** Top panel: Planck functions (normalized flux density, W/m$^2$/µm) for 215 K (Mars thermal emission now, red) and 6000 K (insolation, blue). For context, the $CO_2$ band is overlain (gray shading) at 12-16 µm. Middle panel: Solid black line corresponds to total extinction, dotted black line to scattering. Lower panel shows scattering asymmetry. Also shown are wavelength dependence of total extinction and asymmetry factor for natural dust assuming a log-gaussian particle size distribution centered on 2.5 µm (48) (red dotted lines).

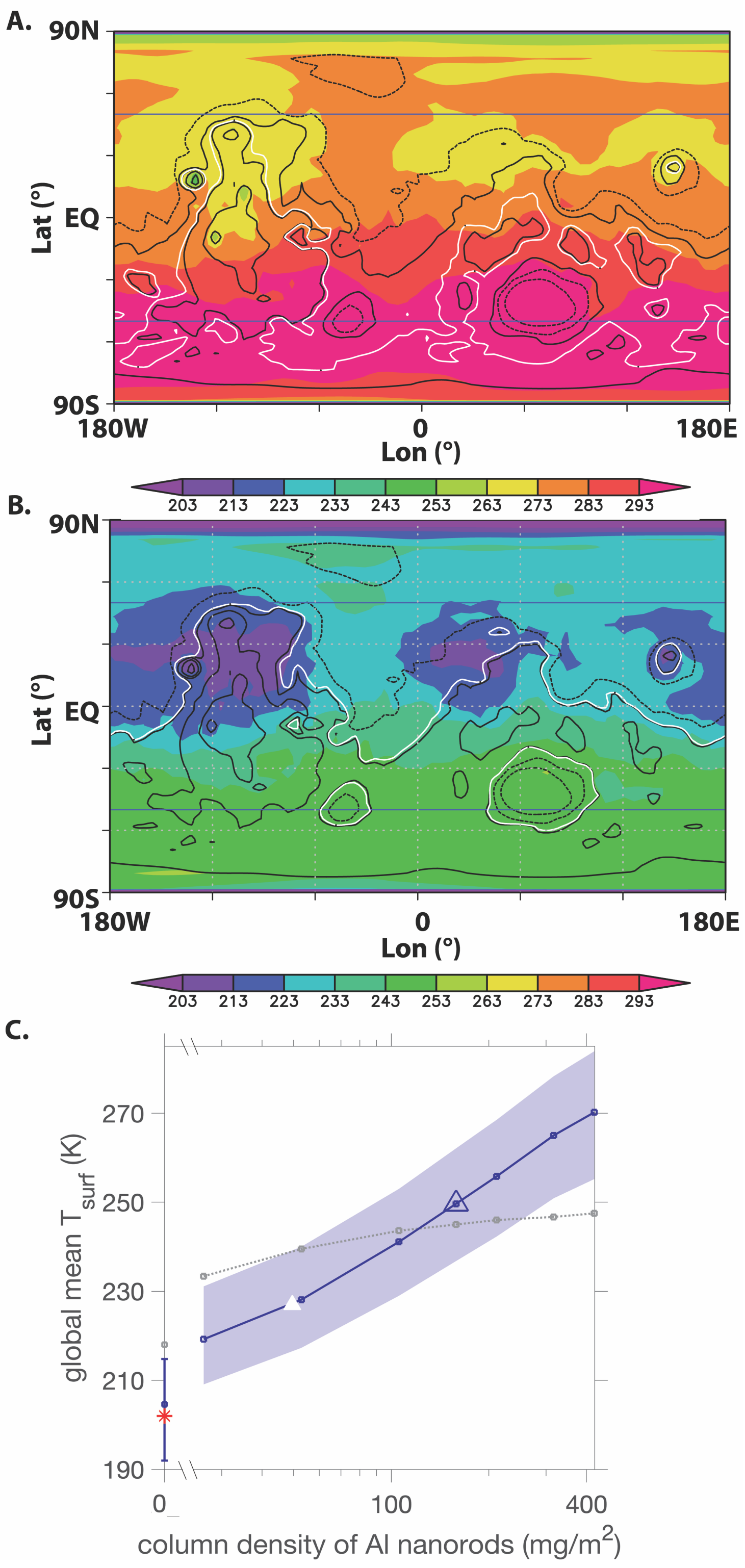

**Fig. 2. 3D model output.** Warm-season temperatures (K) (color shading) on (A) Mars with addition of ~160 mg/m$^2$ of Al nanorods, (B) control case. This corresponds to the average surface temperature during the warmest 36° of solar longitude (~70 days) of the year. White contour corresponds to 610 Pa (~6 mbar) mean pressure level. Black contours correspond to topographic elevation in m (dashed: -5 km and –2 km, solid: 0 km, +2km, and +5 km). Blue lines: approximate latitudinal (equatorward) extent of ice at <1 m depths. Results do not include $CO_2$ outgassing from within polar ice, which would cause further warming. (C) Dependence of planet-averaged surface warming on Al-nanorod column mass. The blue envelope corresponds to the modeled seasonal range in global mean $T_{surf}$. Gray corresponds to 1-D results (see text for details). Blue line corresponds to 3-D results. Blue triangle corresponds to panel (a), and white triangle marks onset of warm-season temperatures above the freezing point of water at 50°S. Symbols on y-axis are temperatures for the no-nanorod case, with the red asterisk corresponding to observed Mars value. Additional details including comparison to Fe nanorods are shown in Figs. S10-S19.

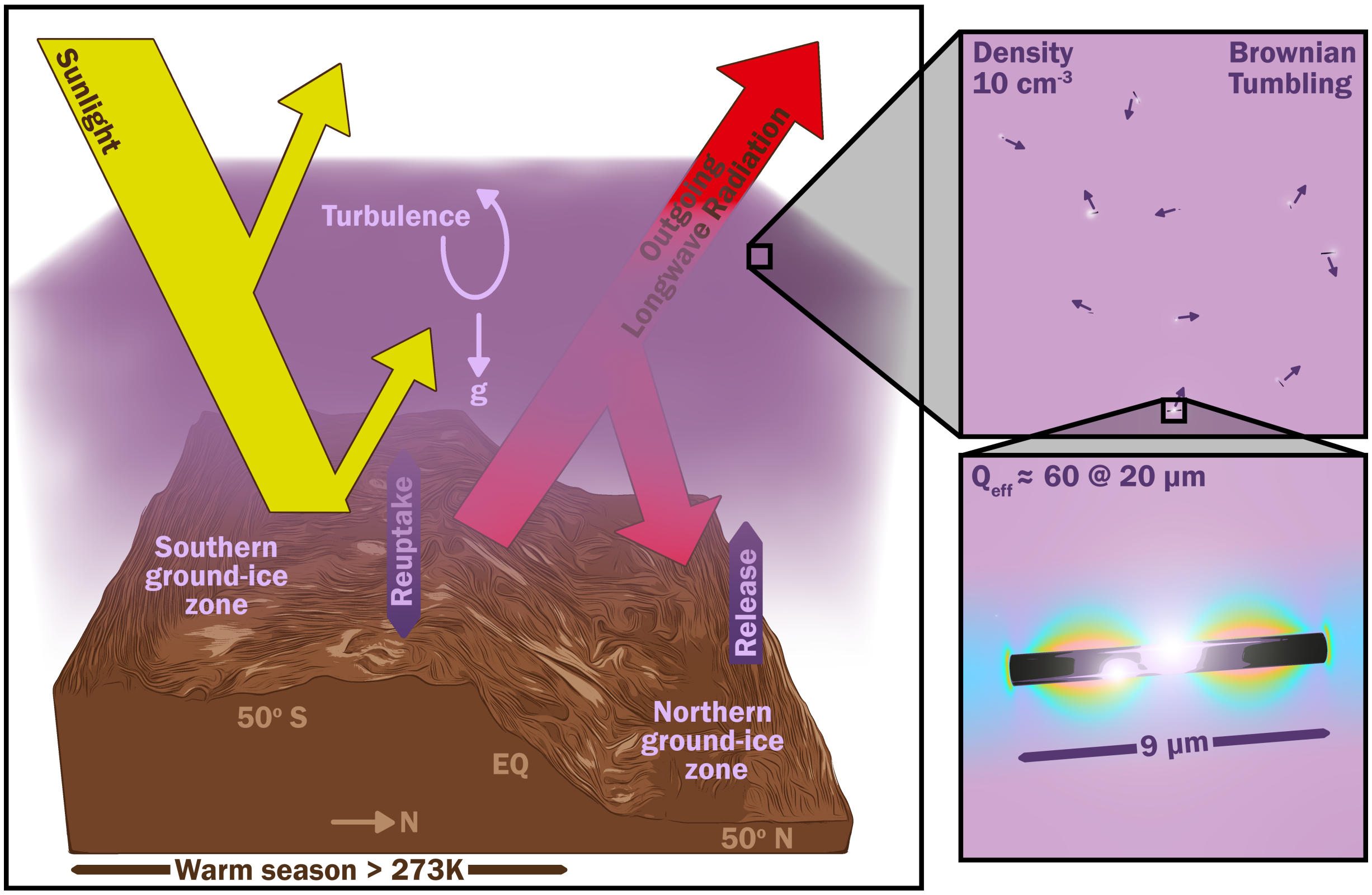

**Fig. 3. Graphical summary.** This figure shows a graphical summary of the proposed warming method. Figure credit: Aaron M. Geller, Northwestern, CIERA + IT-RCDS.

# Supplementary Materials for

## Feasibility of keeping Mars warm with nanoparticles

Samaneh Ansari *et al.*

*Corresponding author: Edwin S. Kite. Email: kite@uchicago.edu

**This PDF file includes:**

Supplementary Text
Figs. S1 to S19
Tables S1 to S3

**Supplementary Text.**

1. Calculation of optical properties of nanorods.

Simple calculations (e.g., ref. 17) suggest that 9 μm long conductive nanorods with a 60:1 aspect ratio would have a strong and broad extinction in the ~22 μm spectral window. To test this, we carried out finite-difference time domain simulations (FDTD: 3D Electromagnetic Simulator). First, we verified that the FDTD simulations reproduce Mie theory for water ice spheres (49) (Fig. S1). The nanorod FDTD simulations use a pulse of light whose interaction with the simulated nanoparticle is Fourier-decomposed to obtain the $\lambda$ -dependence of the absorption and scattering cross sections as well as the scattering asymmetry. The representation of the angular distribution of scattered light by a single parameter, the scattering asymmetry, is standard in climate modeling (e.g., 28). 75 wavelengths, approximately log-uniformly spaced from 0.24-55 μm, are obtained (Table S1). We used refractive indices for Fe from Ref. 50, and for Al from Ref. 51. Ref. 52 suggest that Fe refractive indices obtained prior to their own work might be affected by Fe-oxidation, but as Mars' atmosphere is ~0.1% $O_2$ some degree of oxidation is inevitable, so this is acceptable. Implementation involved combining three different simulations for different $\lambda$ ranges, for computational feasibility. 75 simulations were carried out for the 9 μm long nanorod, corresponding to the product of 5 orientations in θ, 5 orientations in $\varphi$ (0°, 30°, 45°, 60°, and 90°, for both angles), and three $\lambda$ ranges. Here, θ corresponds to rotations in the E-H plane and the other rotation is termed $\varphi$ (k-E plane, symmetric and similar to k-H plane) (Fig. S2). Far-field methods were employed to obtain the scattering phase function. We anticipate that actual nanorods will have circular cross-sections. Because of computational limitations associated with the FDTD approach, we model a nanorod with square cross-section, but find that switching between circular and square cross-sections makes negligible difference to the calculated optical properties (Fig. S3).

As expected, the simulated nanorods showed strong, broad absorptions near 22 μm (Fig. 1), supporting their suitability for Mars warming. This is consistent with previous work on Ag nanorods (53). The wavelength of the absorption peak is slightly longer than double the rod length, due to plasma effects (54). In order to average over orientations (Fig. S4), rod orientation was uniformly sampled on the sphere relative to the incident electric field vector (~1000 samples between 0° and 90° for both angles) and optical properties interpolated using a spline in the grid of 25 computed orientations for each $\lambda$. The orientations of nanorods in the atmosphere are assumed to be random and uniform, hence averaging over a spherically equidistant grid of orientations would result in proper orientation-averaged characteristics. We found (minor) infrared back-scattering in these simulations. A sensitivity test using the 3D model for Fe nanorods showed that this choice makes little difference (~0.3 K) to the calculated temperatures.

To check the interpolation accuracy, we did check runs on multiple intermediate points (Fig. S5). When the rod is viewed nearly end-on, the interpolation results grow less accurate. However, end-on geometries are infrequently encountered, and at the higher angles, the scattering and absorption cross-sections are substantially smaller, so the effect of these points on the overall orientation average is insignificant. To check the simulation accuracy, we did a check run increasing the resolution across the width of the nanorod from 8 simulation mesh points (20 nm spacing) to 10 simulation mesh points (16 nm spacing), finding negligible differences (Fig. S6).

We also did calculations for 7.5 μm-long Fe nanorods with cross-section 0.08 μm × 0.08 μm, finding peak extinction around 20 μm (Fig. S7), and less climate warming per nanorod, but more climate warming per kg of nanorods. An optimal warming approach might use a mix of rods of different lengths. Moreover, we find that the total extinction (absorption plus scattering) for Al nanorods is the same as for Fe nanorods (Fig. S8), and the corresponding warming is also about the same (Table S2), even though the density of Al is three times less than that of Fe. Fe, Al, and Mg are all present at >4 wt% level in Mars soil. We focus here on Al and Fe because optical properties are available over the full wavelength range of interest (other metals are worthy of future investigation). This suggests that further research might yield further improvements in the effectiveness of warming.

2. Calculation of the surface-warming effect of the nanorods using 1D climate model.

Our single-column radiative-convective climate model (RCM) subdivides atmospheres into multiple vertical log-layers (we implement 201 layers here) that extend from the ground to the top of the atmosphere ($1 \times 10^{-5}$ bar here) (e.g. 25). The RCM has 55 infrared and 38 solar spectral intervals (55). The model applies a standard moist convective adjustment (e.g. 56). Should tropospheric radiative lapse rates exceed their moist adiabatic values, the model relaxes to a moist $H_2O$ adiabat at high temperatures or to a moist $CO_2$ adiabat when temperatures are low enough for $CO_2$ to condense. The RCM implements a standard solar spectrum (57). For present Mars (solar flux = 585 W/m$^2$), we assume a typical stratospheric temperature of 155 K and a surface albedo of 0.22 (e.g. 24). The atmospheric pressure is a Mars-like 650 Pa with an assumed acceleration due to gravity of 3.73 m/s$^2$. Although we prescribe a tropospheric relative humidity of 50%, our results are insensitive to this parameter. The baseline mean surface temperature of the resultant pure $CO_2$ atmosphere (without nanorods) is 218 K. For comparison, Mars' observed global mean surface temperature is 202 K (10). The 1D model runs warmer in the 6 mbar no-nanorods case mainly because it lacks day-night and equator-pole temperature contrast, and because it lacks clouds, dust and topography (10). With an assumed surface albedo of 0.22 and solar flux of 585 W/m$^2$, the predicted equilibrium temperature ($T_{eq}$) would be ~212 K (assuming a perfect blackbody). This 212 K $T_{eq}$ value is the minimum surface temperature for a 1D model with no greenhouse effect. However, our 1D model predicts 6 K greenhouse effect (consistent with data and with other models), raising the mean surface temperature to ~218 K (e.g. 58). In summary, the warm 1D model output gives a useful benchmark comparison on the importance of 3D factors like dust, clouds and topography on modeling predictions in comparison with that from the nanorods.

One of the reasons for the differences in temperature response between the models is the baseline surface temperature, which is higher in the 1D model (for reasons explained above). The main cause, however, is differences in the vertical temperature profiles across models (e.g., Fig. S18). For instance, lapse rates differ slightly (the top of the cloud deck is at a slightly higher pressure level, or closer to the ground, in the 3D model) and the 3D model has a near-surface temperature inversion which is absent in the 1D code (Fig. S18). The 3D model also has a larger effective convective region. Tests confirm that the 1D and 3D results agree more closely when the 3D profile is imposed on the 1D model. For example, at $\tau_{Fe}$ = 0.25, imposing the same mean surface temperature in the 1D model as is obtained for the 3D model (~225.5 K) yields a top-of atmosphere (TOA) net outgoing longwave radiation/net incident stellar flux ratio of ~0.69 (a value equal to 1 corresponds to a TOA radiative equilibrium). In other words, the 1D model "wants" to warm. However, when the 3D model vertical temperature profile was applied instead (at the same surface temperature), this ratio had increased to 0.86 (removing ~60% of the difference from 1). In other words, most of the difference between the 3D model and the 1D model can be explained by differences in the vertical thermal profile. This suggests that even across 3D models, differences in convective schemes (and temperature profiles) could produce some spread in the results, motivating investigation of engineered warming of Mars using different 3D schemes. Despite these numerical differences, the atmospheric response to nanorod addition in both the 1D and 3D models is qualitatively very similar. In both cases, nanoparticle warming provides a greenhouse effect >5,000× greater than the current state of the art.

Following Refs. 24-25, we calculate the wavenumber-dependent optical depths ($\tau$) for the nanorods from the following expression:

$$\tau = 3\, Q_{eff}\, NRC\, \Delta z \,/\, (4\, r\, \rho) \quad (2)$$

Here, *NRC* is the nanorod content (g/m$^3$) and $\Delta z$ is the path length. This equation differs from (1) only in that the optical depth is integrated over all nanorod heights and across all wavenumbers, not just in the spectral window. The nanorods are well-mixed throughout the atmosphere (up to 35 km) above the imposed 500 Pa nanorod-layer base and so *NRC* was assumed to scale linearly with the local pressure. We computed the radius of the sphere with the equivalent nanorod volume, which yields an effective nanorod particle radius of ~0.38 μm for the 9 μm nanorods.

Within the 1D model framework, we implement a simple procedure to calculate nanorod warming. For each assumed nanorod optical depth, we find the surface temperature that yields stratospheric energy balance (i.e. the net outgoing and net incoming fluxes must balance each other) (Figs. S16-S17).

3. Calculation of the surface-warming effect of the nanorods using 3D climate model.

The FDTD output is interpolated to 2000 log-uniformly spaced $\lambda$'s. We set optical properties at $\lambda$ >55 μm equal to those at 55 μm (Fig. S9). This approximation is acceptable, because extinction (W/m$^2$/μm) is minor at such long $\lambda$. MarsWRF uses a two-stream radiation code (e.g. 22, 59). Radiative transfer calculations include both gas (for this work, $CO_2$) and aerosol (for this work, natural dust, nanorods, and $CO_2$ ice) radiative effects. This implementation of MarsWRF uses Planck-weighted averaging to bin down high-spectral-resolution optical properties. The blackbody temperature used for the Planck-weighted averaging is 6000 K for solar bands (6 bins from 0.24-4.5 μm, with bin edges at 0.24, 0.40, 0.8, 1.31, 1.86, 2.48, 3.24, and 4.5 μm), and 215 K for the thermal infrared bands (6 bins from 4.5-1000 μm, with bin edges at 4.5, 8.0, 12.0, 14.1, 16.0, 24.0, 60, and 1000 μm) (Fig. S9). Thermal equilibrium between gas and dust is assumed. We neglect radiation pressure, magnetic effects, quantum size effects, and temperature dependence of the optical constants.

In the fixed-cloud runs, present-day Mars values are used for orbital parameters, surface albedo, and surface thermal inertia. Mars' southern summer solstice occurs near perihelion, so (by Kepler's second law) northern summer is long and relatively cool, and southern summer is short and warm relative to annual average (Figs. S10-S12). Water vapor abundance in the atmosphere is fixed to zero, which is conservative in that water vapor radiative feedback would increase warming. A prescribed natural dust aerosol distribution is imposed corresponding to the Mars Climate Database MGS-dust-scenario (60), giving an average dust optical depth of 0.20 (peaking at 0.43 during southern summer). Natural dust aerosol warms the atmosphere but lowers daytime surface temperature. The artificial-aerosol layer is parameterized by a layer-base pressure (in Pa), a layer vertical thickness (in units of model levels), and the nanorod orientation-averaged optical depth at a reference $\lambda$ (0.67 μm). Because the model levels are specified in $\eta$ coordinates (where $\eta = P/P_{surf}$, where the model-top pressure has been subtracted from both pressures), the aerosol layer is physically more compact over the poles because low temperatures compact the air column. We refer to the altitude away from the poles below which 95% of the particles are contained as the "aerosol layer top height". The nanorod mixing ratio is assumed to be uniform within the nanorod layer (Fig. S13). We prescribe a layer-base level of 500 Pa and vary the layer thickness. This approach gives a maximum in artificial-aerosol opacity above the surface, which is also observed for natural Mars dust aerosol for some seasons (61-63). However, natural dust on Mars persists all the way to the surface (63) which has average pressure ~600 Pa. This has no effect on our conclusions, as results are only weakly sensitive to the vertical distribution of simulated aerosol - for example, in Table S3, the "Thinner cloud (200 Pa base, 8 levels)" run differs from the reference case with 500 Pa base by only 0.3 K in average temperature. Buffering by latent heat implies that melting either ice cap to form seas would take at least centuries.

Settling for a 0.1 μm-radius dust particle for 1 scale height at Mars surface pressure takes 3 years (Murphy 1990). Extrapolation to Al nanorods with radius 0.04-0.08 μm suggests settling times that are similar or greater. Increase of Mars' atmospheric $pCO_2$ under warming will decrease the Knudsen number and hinder dust settling. One-pass settling rates are much slower than that of natural Mars dust, and effective particle lifetime will be longer due to re-entrainment.

The runs presented here use a horizontal resolution of 5.625° × 3.75°, corresponding to a grid of 64 points in longitude × 48 points in latitude. A 40-layer vertical grid is used, using a modified-$\eta$ (terrain-following) coordinate ($\eta = P/P_{surf}$, where the model-top pressure has been subtracted from both pressures). The dynamical timestep varies between runs but is never longer than 1-min. The planetary boundary layer scheme is based on that in Ref. 64 (Medium-Range Forecast model scheme), with Mars implementation similar to that in Ref. 20, and the surface layer uses a Monin-Obukhov scheme. Runs are initialized from a cold state and continue until the simulated annual seasonal cycle is highly repeatable from year to year. We found that runs require only a single year of spin-up adjustment, as modern Mars, lacking seas, has low effective thermal inertia. Output is reported from the third year of each model run. Mean wind speed ~30 m above the surface increases from 8 m/s in the no-nanorods case to 13 m/s in the Fig. 2a (with-nanorods) case, which would stir up more dust (and nanorods). Daytime near-surface turbulence on Mars is sufficient to loft dust (and therefore nanorods) released at 10-100m altitude at all latitudes for P > 6 mbar. Sensitivity tests adjusting model resolution, and varying nanorod distribution and other parameters are listed in Table S3.

For the no-nanorods case (Fig. 2b), the global and annual average shortwave radiation reaching the surface is 133.1 W/m$^2$ (119.7 W/m$^2$ direct beam and the remainder diffuse/scattered), of which 104.2 W/m$^2$ is absorbed with the rest being reflected. The spatially- and time-averaged surface albedo, including ice, is 0.237. The global/annual average longwave radiation at the surface (greenhouse effect) is 23.9 W/m$^2$. The global/annual average longwave emission from the surface is 125.4 W/m$^2$. The mismatch of -2.7 W/m$^2$ is caused by limited output sampling, fluxes into the surface (e.g. $CO_2$ condensation, conduction), and model imprecision. For the with-nanorods case shown in Fig. 2a, the global and annual average shortwave radiation reaching the surface is 89.2 W/m$^2$ (41.4 W/m$^2$ direct beam), of which 70.6 W/m$^2$ is absorbed with the rest being reflected. The spatially- and time-averaged surface albedo, including ice, is 0.208. A greater fraction of sunlight is absorbed due to reduction in reflective seasonal $CO_2$ ice. The global/annual average longwave radiation at the surface (greenhouse effect) is 186.2 W/m$^2$. The global/annual average longwave emission from the surface is 254.3 W/m$^2$. The mismatch of -2.5 W/m$^2$ (i.e., surface temperature under-stated relative to radiative fluxes) is caused by limited output sampling, fluxes into the surface (e.g. $CO_2$ condensation, conduction), and model imprecision.

The runs shown in Fig. 2c correspond to optical depths at $\lambda$ = 0.67 μm of {0, 0.125, 0.25, 0.5, 0.75, 1, 1.5, 2}. Details for selected values are given in Fig. S14. The conversion factor is 545 mg/m$^2$ for Fe and 212 mg/m$^2$ for Al for one optical depth. Optical depths in the spectral windows are much greater than at 0.67 μm (Fig. 1). Output was sampled at 260 equally spaced intervals per Mars year. This provides (deliberately aliased) sampling of the day-night cycle during each season at each longitude and latitude. A sensitivity test using 1300 output steps showed negligible ($\leq$0.1 K) differences in both annual-average and seasonal warming (Table S3). Mars' atmosphere is thin, with low thermal inertia and limited ability to transport heat laterally, and the tests with artificial imposed patchy or unsteady nanorod distribution show that the corresponding warming is sharply confined in space and time. This suggests that it might be possible to enhance warming at preferred latitudes and seasons. The atmosphere and surface temperature quickly responds to the radiative forcing at the release site/time, and after injection ceases, spreading of the particles away from the release site ensures that the warming at the release site soon decreases.

4. Possible hazards.

Natural Mars air is unsafe for humans to breathe because it has almost no oxygen (insufficient for deflagration) and also has a high natural concentration of PM 2.5 (Mars mineral aerosol dust). The nanorod density is ~10 μg/m$^3$, which would not substantially alter this situation. A more immediate concern is asbestosis, as humans would bring both natural dust and nanoparticles into settlements via airlocks. One way to mitigate this hazard would be to make nanorods that dissolve or fragment in liquid water.

5. Comparison to previous work.

To our knowledge, the most effective (on a per-unit-mass-in-the-atmosphere basis) Mars warming agent that has been previously been proposed is the "optimal [gas] mix" of ref. 8, which is mostly $C_3F_8$ (molecular mass 188 Da). This gives 37.5 K warming for 1 Pa, which corresponds to 170 kg/m$^2$ × (1 Pa / 650 Pa) × (~188 Da / 44 Da) ≈ 1.1 kg/m$^2$. This warming is about the same as the nanorod case shown in Fig. 2a, which corresponds to ~160 mg/m$^2$ of Al nanorods. Therefore this nanorod loading (using non-optimized nanoparticles) is >5000× more effective than the optimal gas mix.

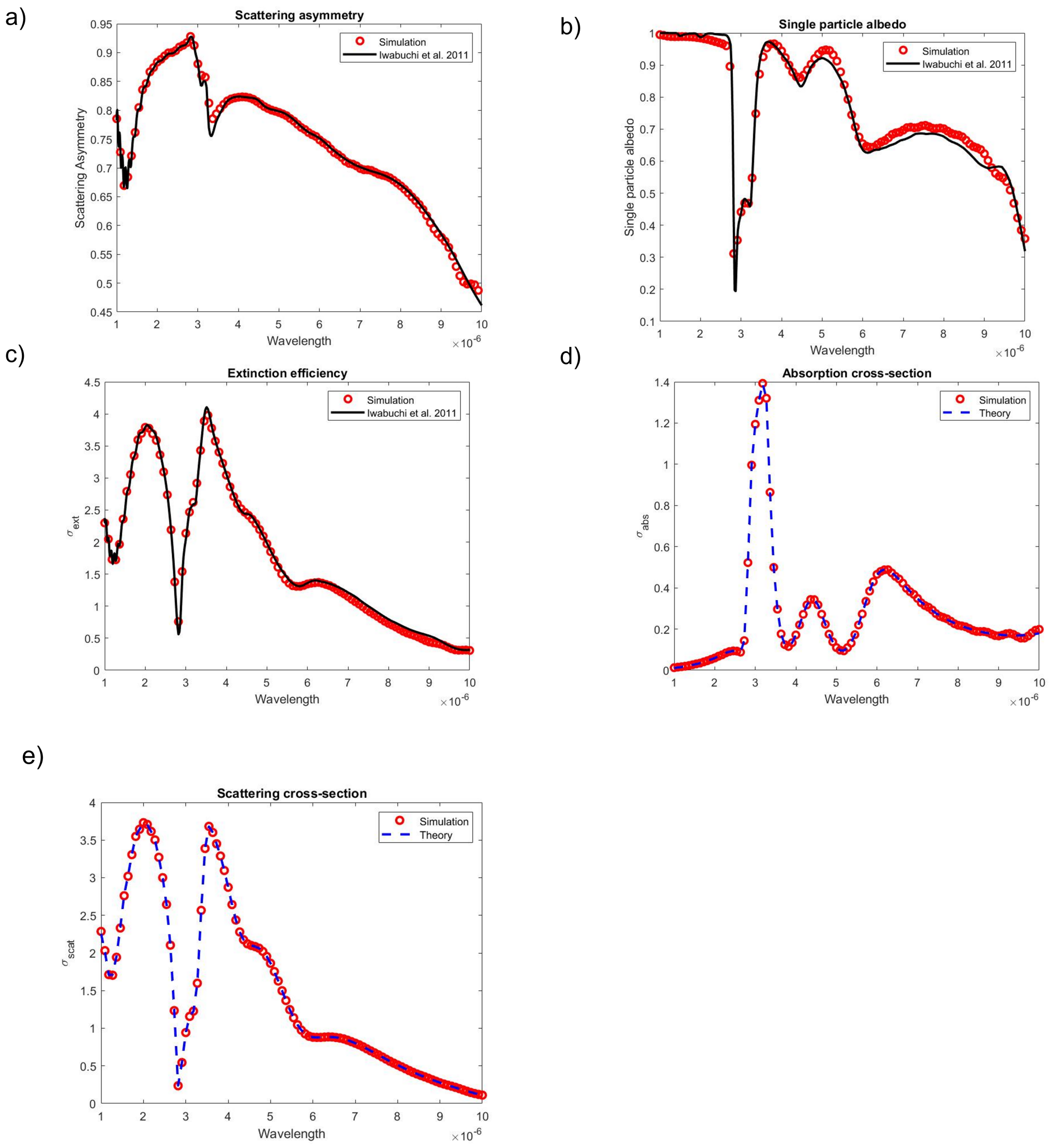

**Fig. S1.** (a,b,c) Verification of calculations by comparison to Mie-theory results for a water-ice sphere (49), showing single particle albedo, scattering asymmetry, and extinction efficiency. (d,e) Shows the comparison of analytical Mie scattering results for an ice sphere with simulation outputs.

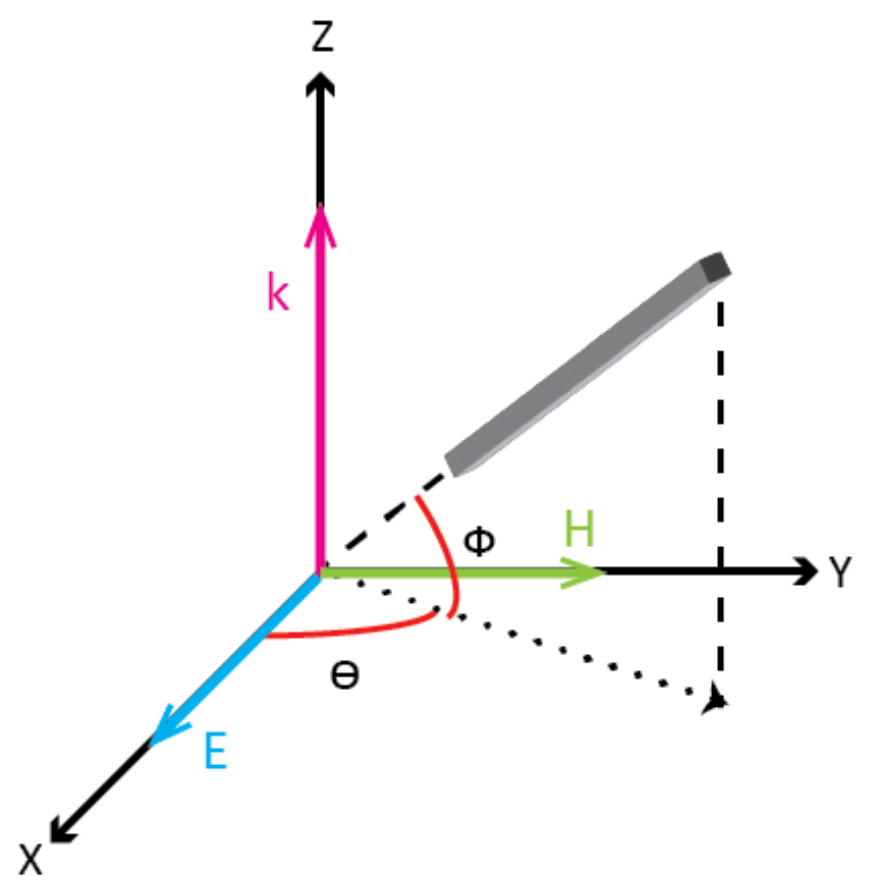

**Fig. S2.** Illustration showing the definition of the angles $\theta$ and $\varphi$ in the FDTD calculation. "k" corresponds to the direction of the propagation of the incident electromagnetic wave. The gray bar corresponds to the nanorod.

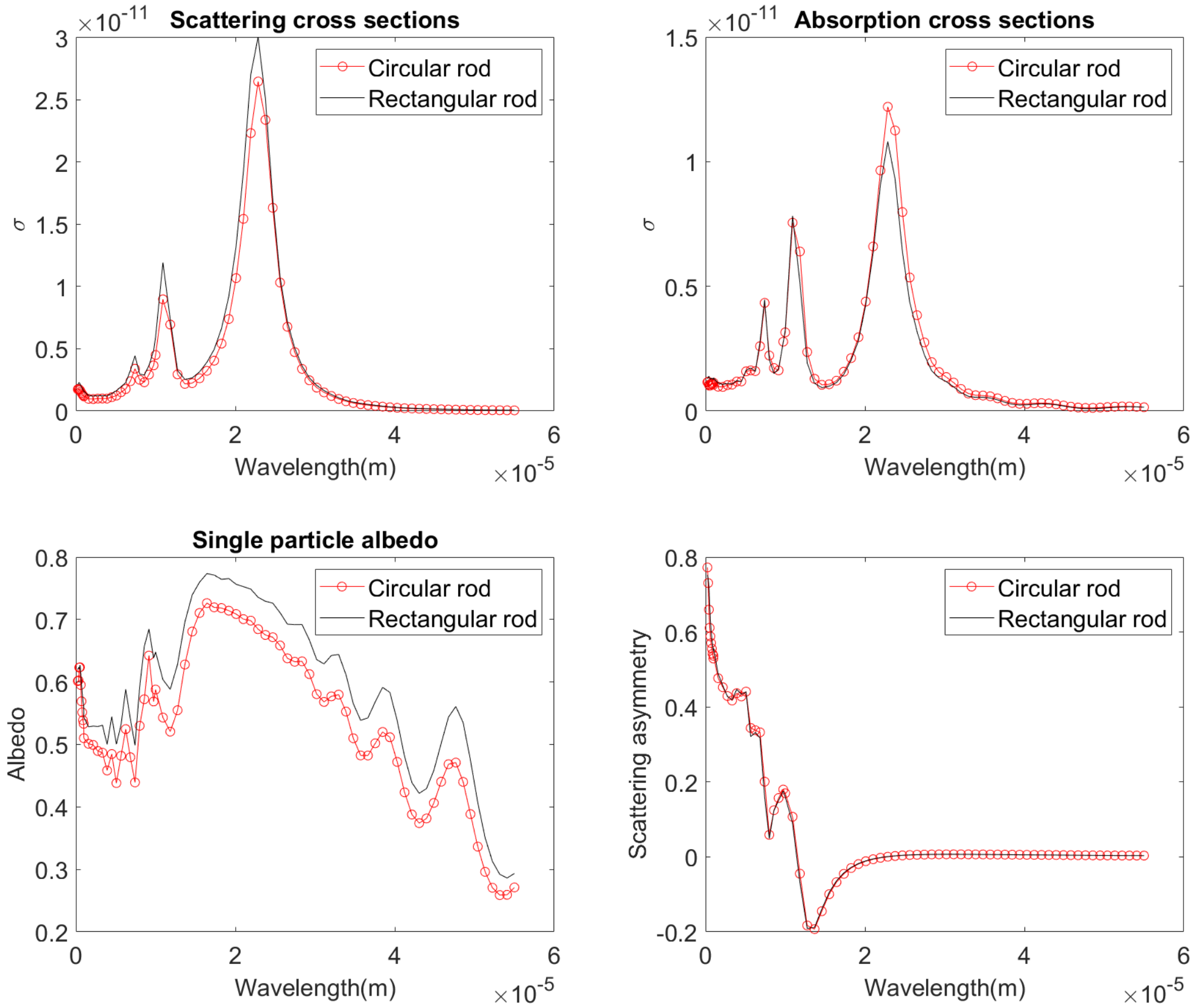

**Fig. S3.** Switching between a square and circular cross-section for the nanorod has only minor effects on the calculated optical properties. Fe nanorod, 9 μm-long, ~60:1 aspect ratio, orientation: $\theta = 45°$, $\varphi = 45°$.

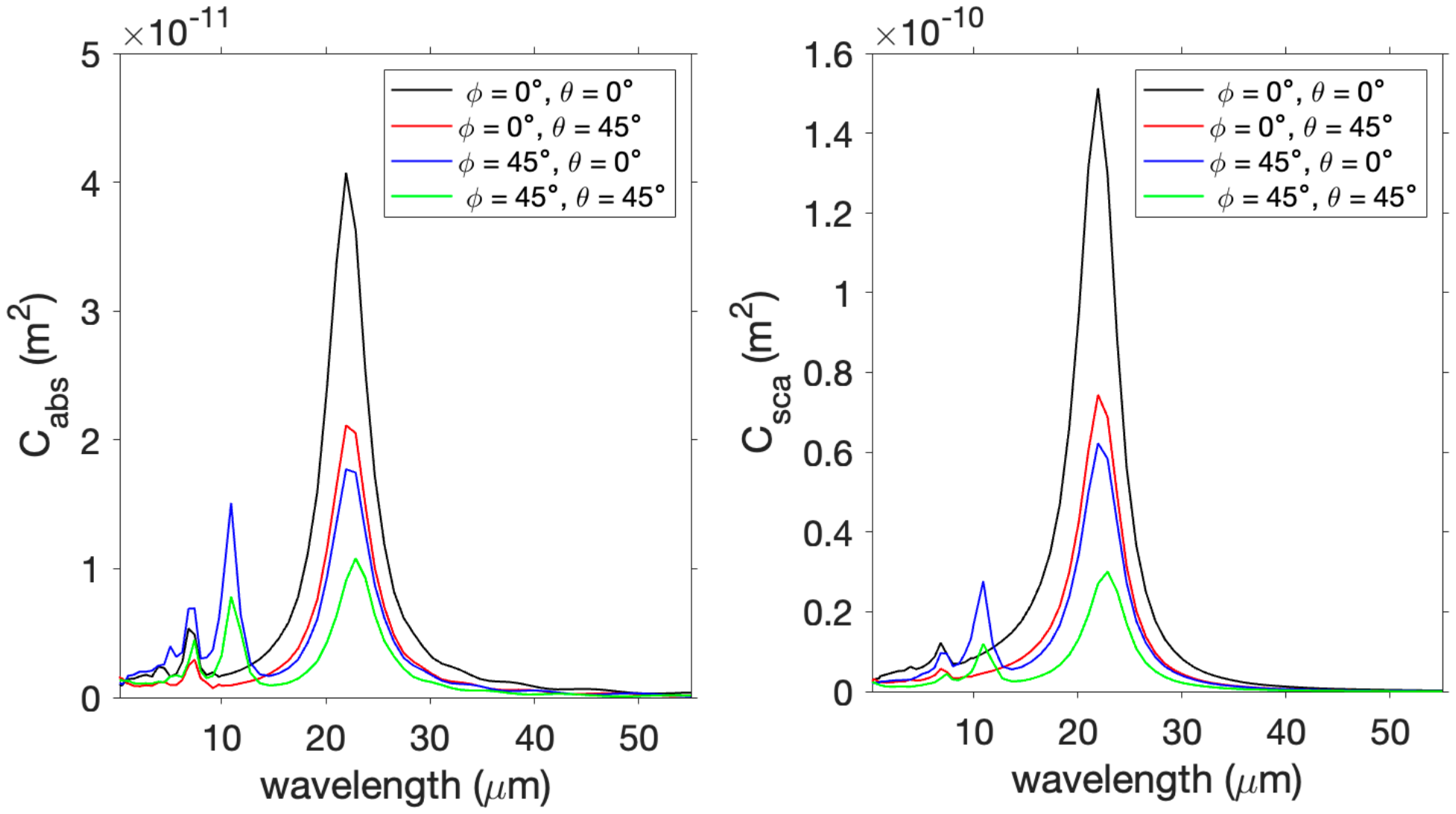

**(a)**

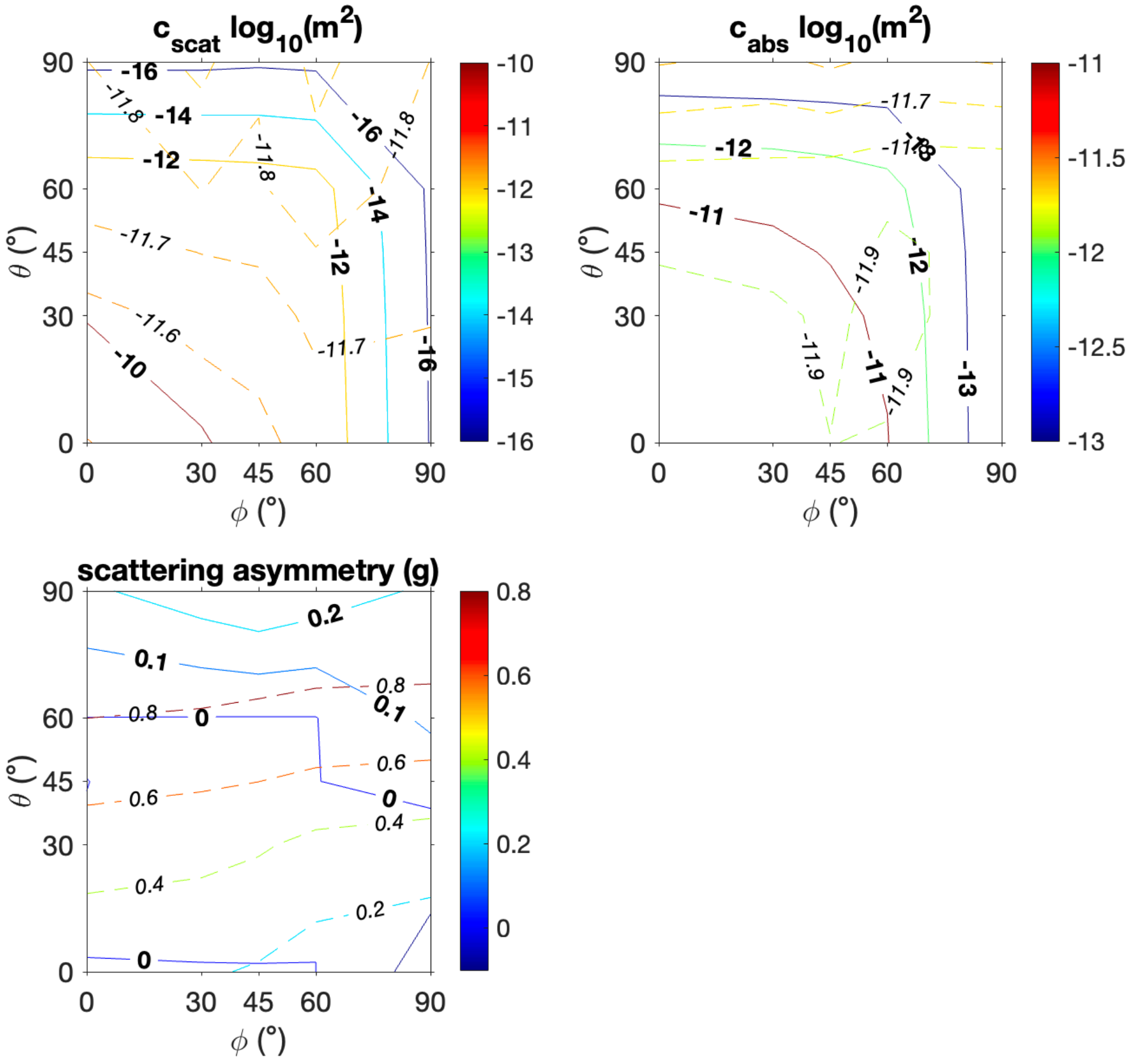

**(b)**

**Fig. S4.** FDTD output figure, showing orientation dependence of the optical properties of a 9 µm-long, ~60:1 aspect ratio Fe nanorod. (a) Cross-sections as a function of wavelength. (b) Cross-sections and scattering asymmetry as a function of orientation for wavelengths 21.9 µm (solid lines, bold labels) and 0.70 µm (dashed lines, italic labels).

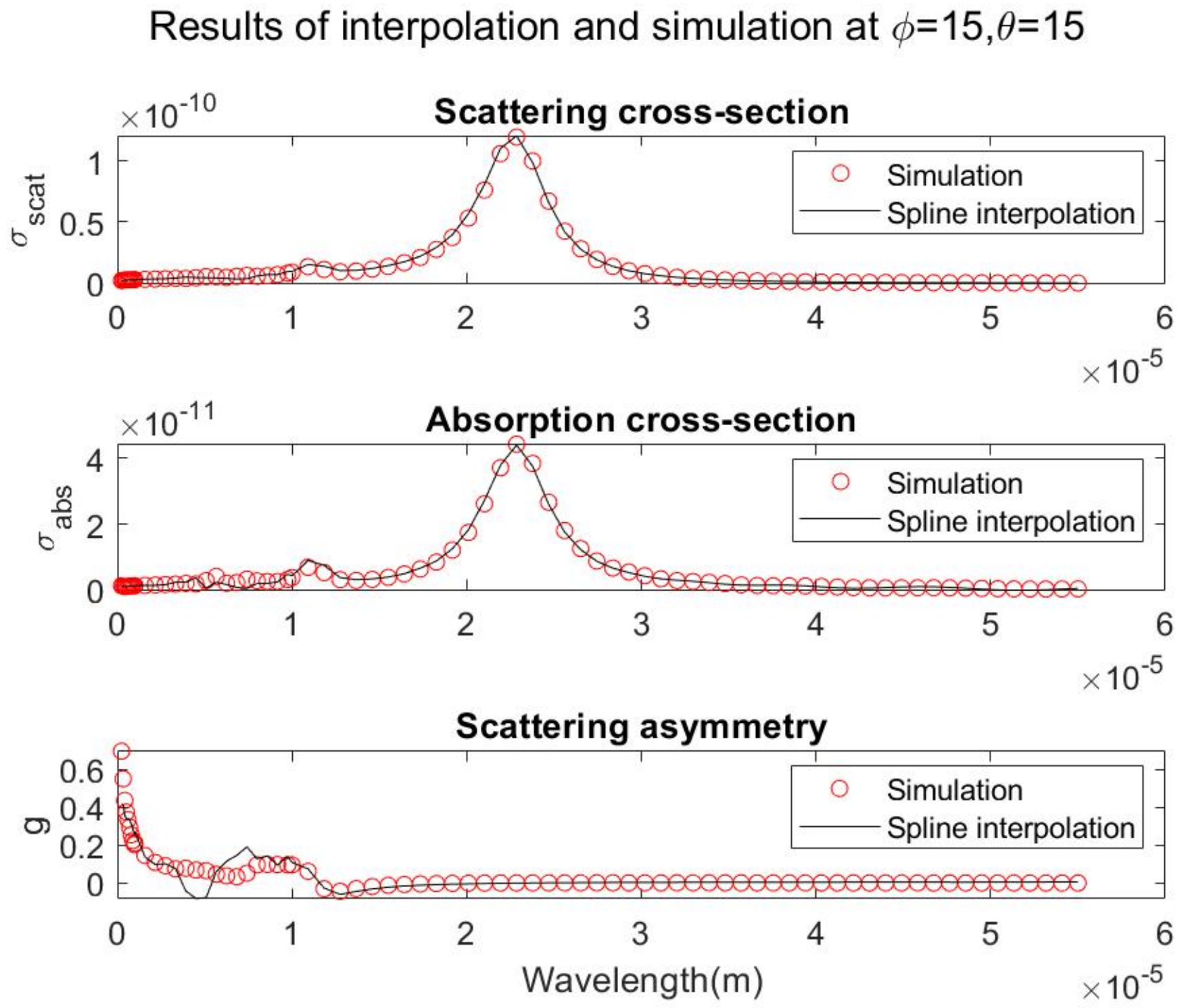
Results of interpolation and simulation at $\phi$=15,$\theta$=15
Scattering cross-section
$\times 10^{-10}$
$\sigma_{scat}$
Simulation
Spline interpolation
Absorption cross-section
$\times 10^{-11}$
$\sigma_{abs}$
Simulation
Spline interpolation
Scattering asymmetry
g
Simulation
Spline interpolation
Wavelength(m)
$\times 10^{-5}$

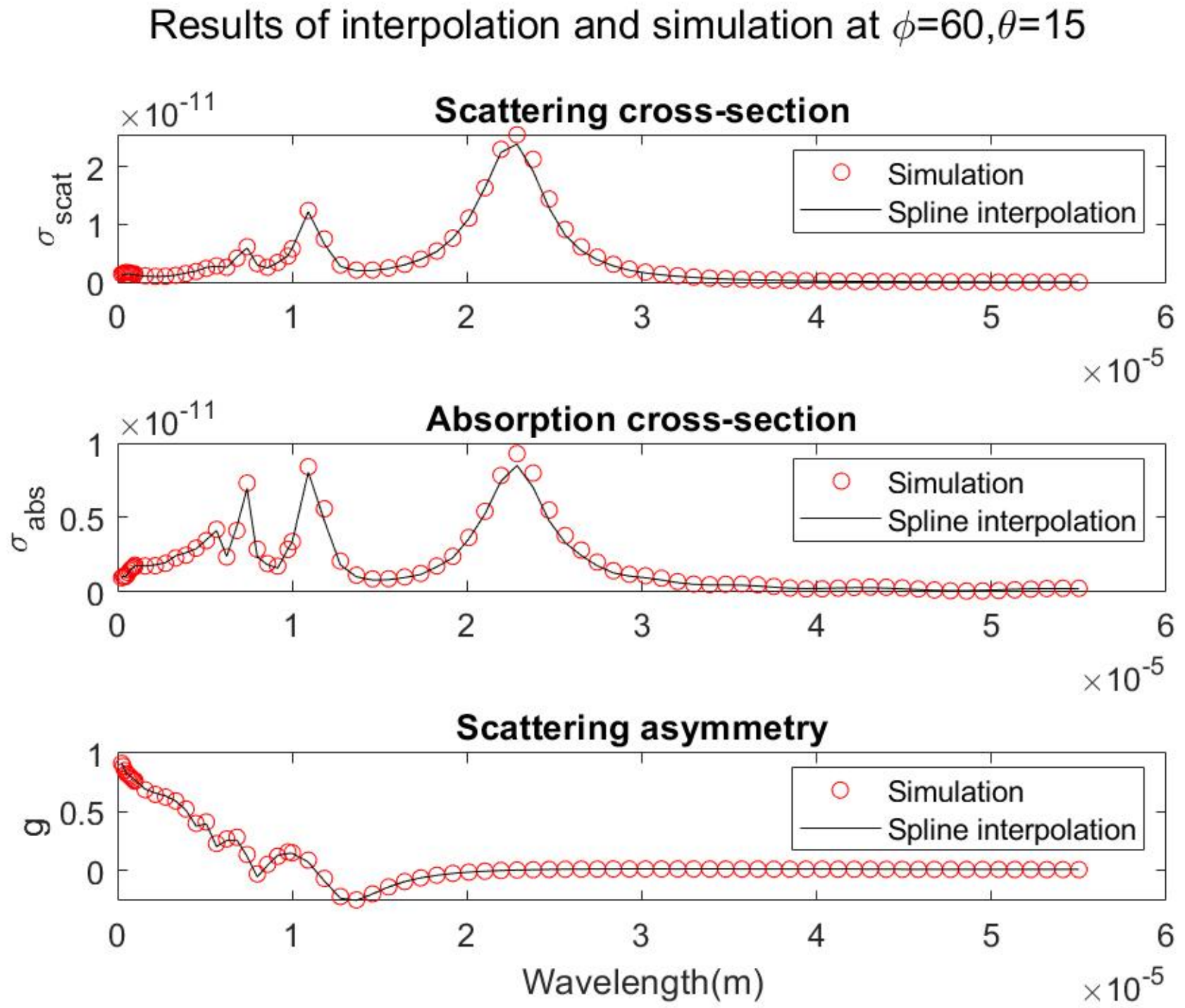
Results of interpolation and simulation at $\phi$=60,$\theta$=15
Scattering cross-section
$\times 10^{-11}$
$\sigma_{scat}$
Simulation
Spline interpolation
Absorption cross-section
$\times 10^{-11}$
$\sigma_{abs}$
Simulation
Spline interpolation
Scattering asymmetry
g
Simulation
Spline interpolation
Wavelength(m)
$\times 10^{-5}$

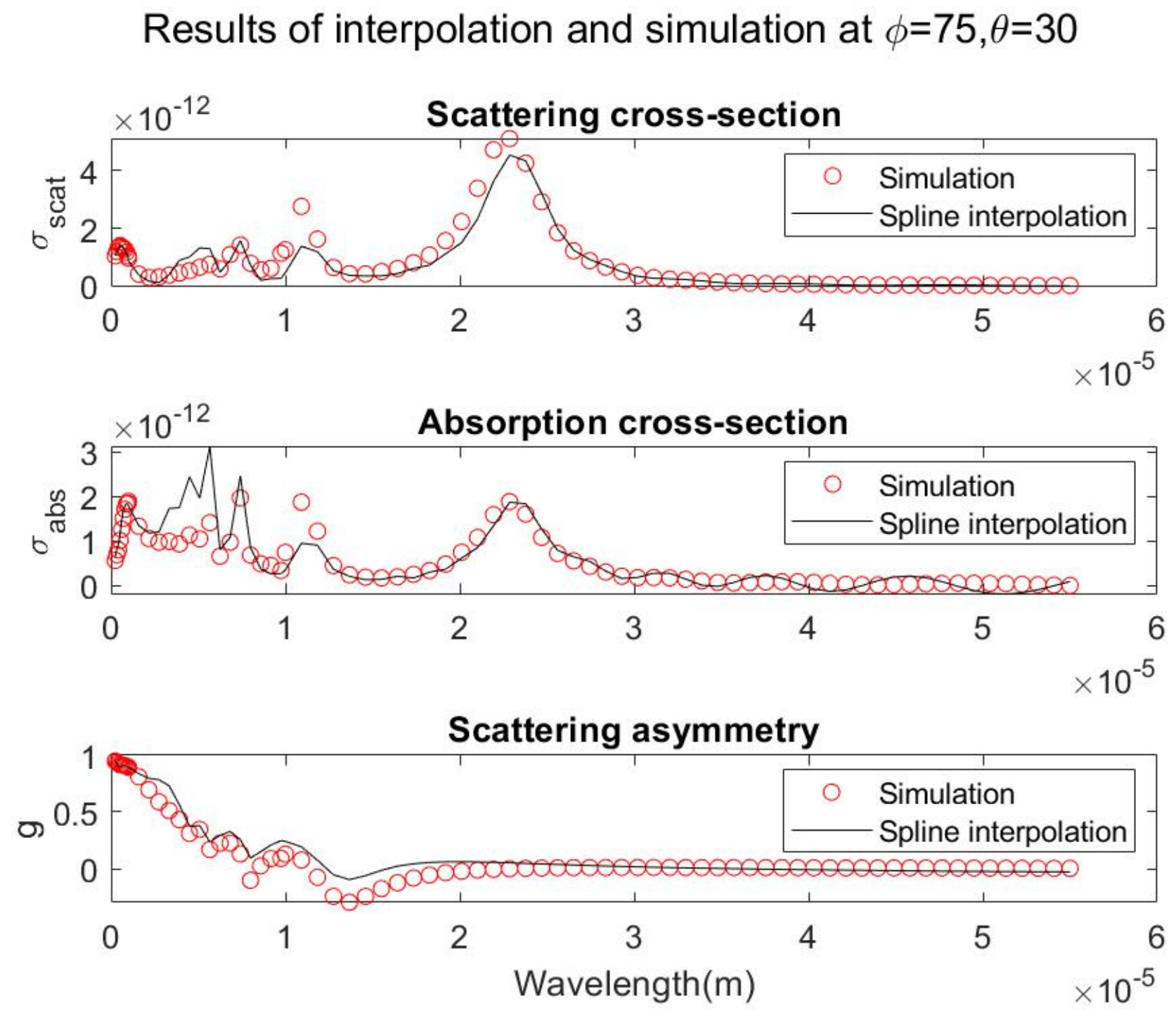

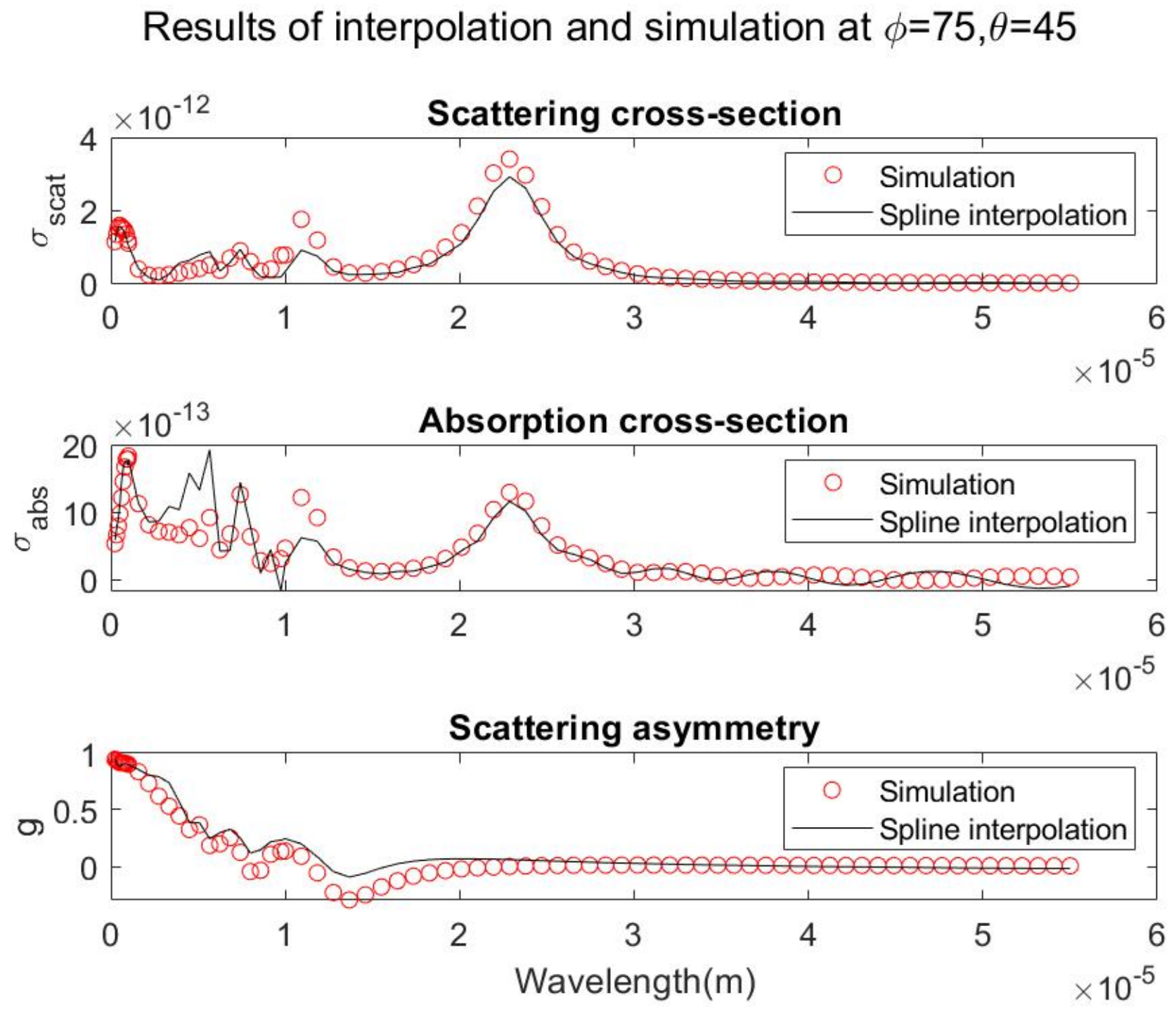

**Fig. S5.** Results of interpolation check. 9 μm-long, ~60:1 aspect ratio Fe nanorod. Units of angles are degrees.

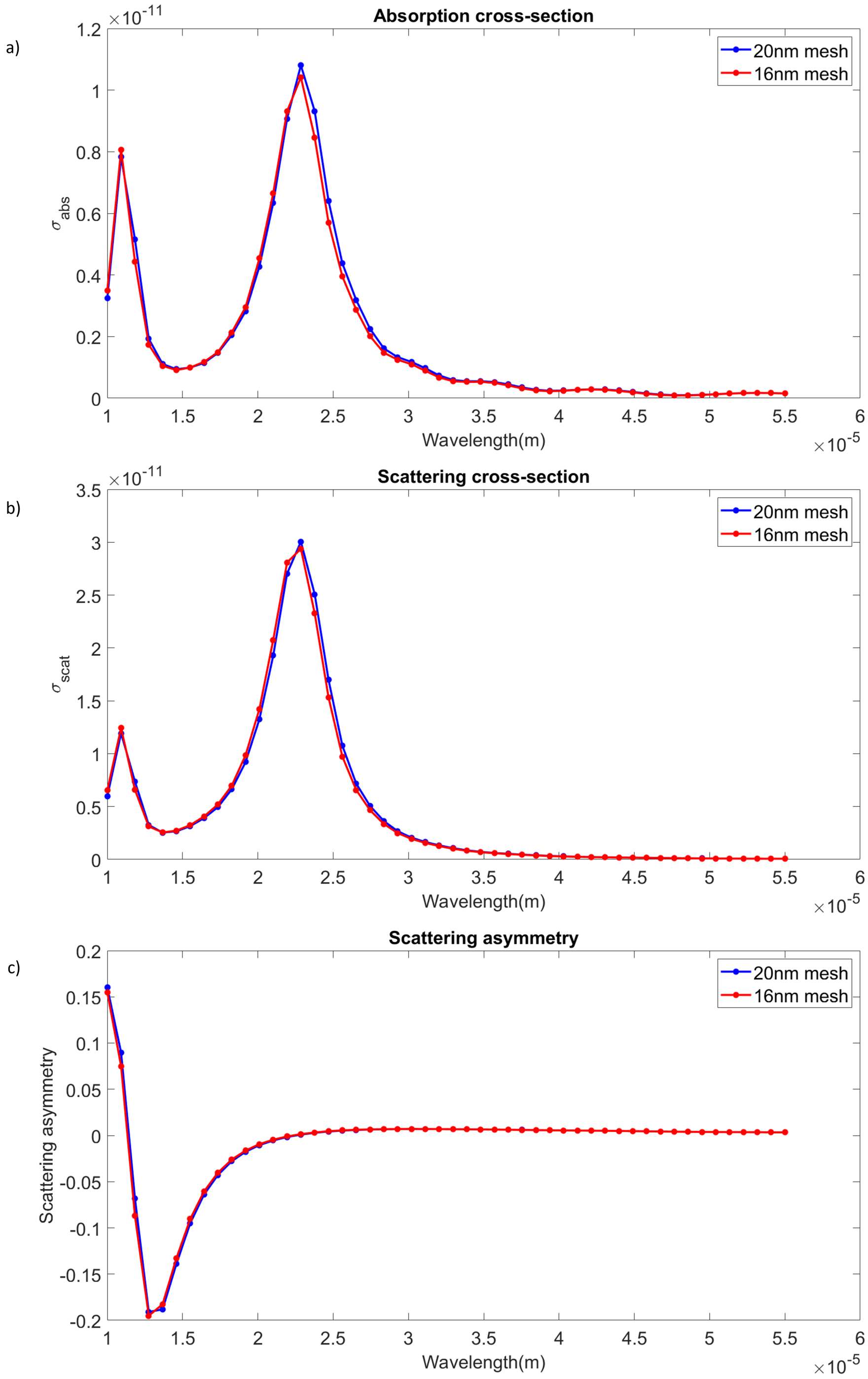

**Fig. S6.** Sensitivity test showing change in optical properties when mesh resolution is increased (decreasing mesh spacing). 9 µm-long, ~60:1 aspect ratio Fe nanorod. Orientation: θ = 45°, φ = 45°.

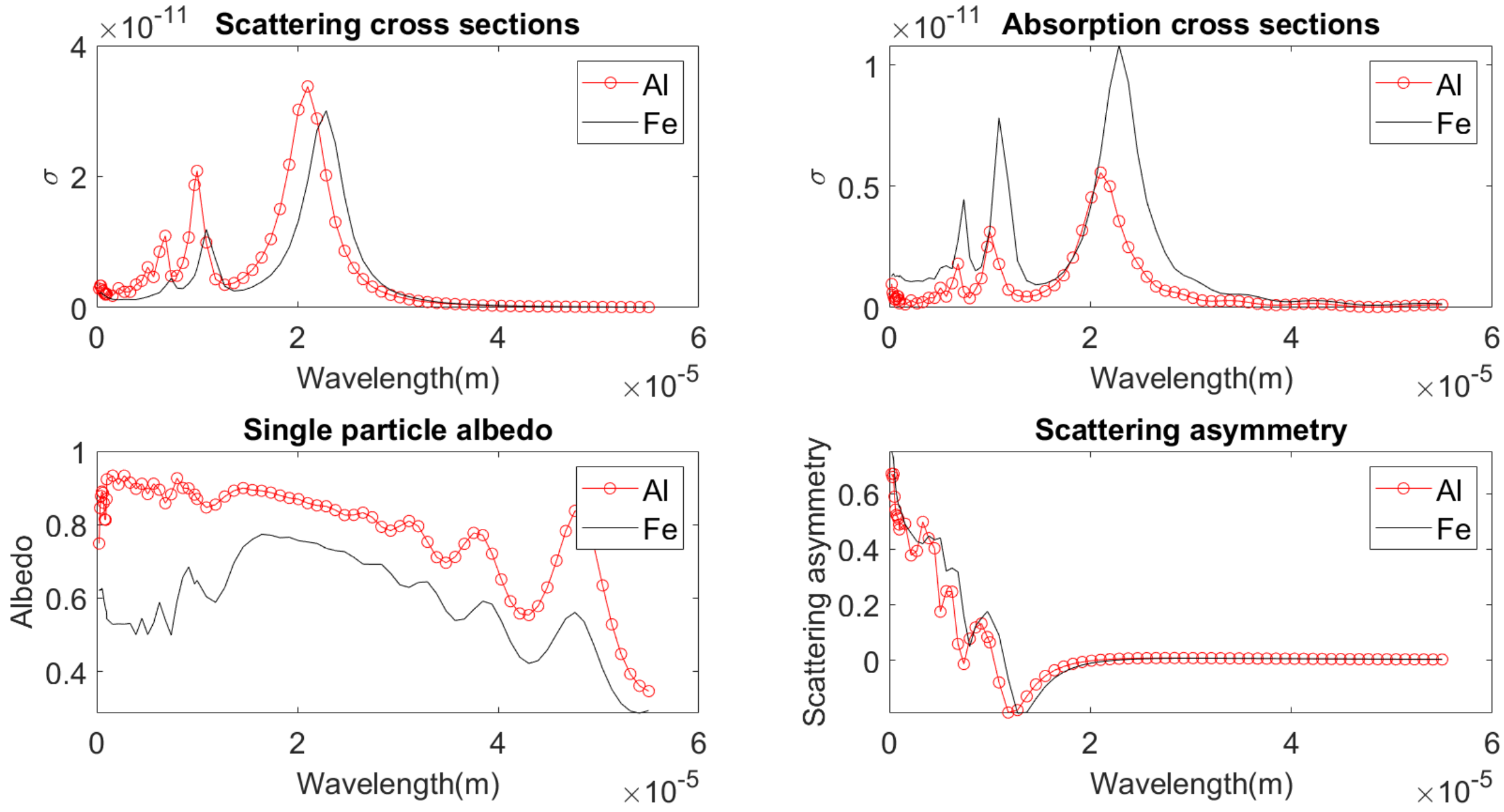

**Fig. S7.** Showing sensitivity of calculated optical properties to changing nanorod composition (Al vs. Fe), for a 9 µm-long, ~60:1 aspect ratio nanorod. The total extinction cross section (scattering plus absorption) and the location of the resonances remains about the same; a small increase in Al nanorod length would be sufficient to closely match the Fe cross-sections. As Al is 3× less dense than Fe, this suggests Al is more effective on a warming-per-unit-mass basis. Conductive-nanorod size and shape, and not composition, are the main controls on simulated nanorod optical properties. Aluminum material properties are obtained from ref. 51. Orientation: $\theta = 45°$, $\varphi = 45°$.

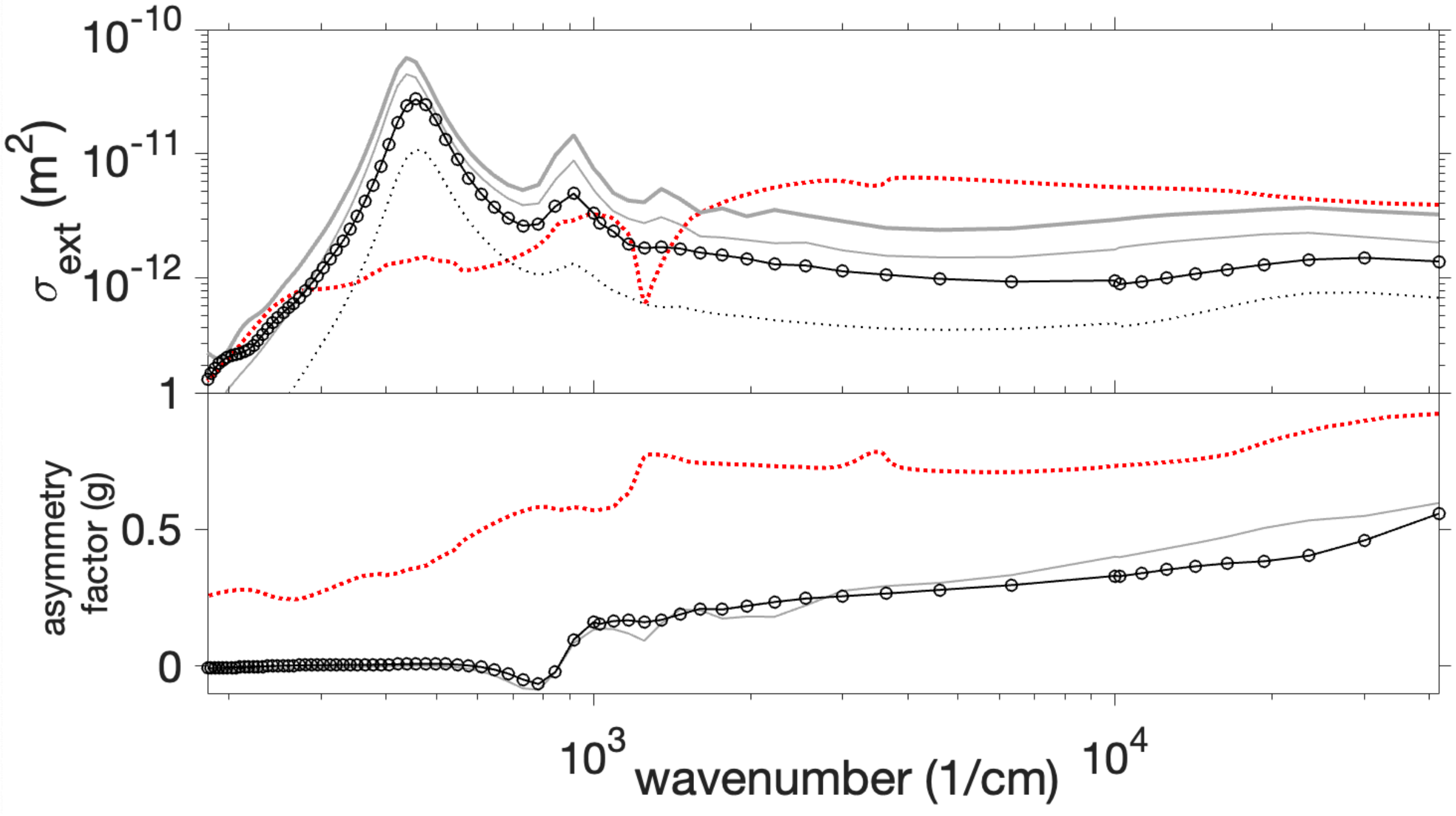

**Fig. S8.** Showing the calculated optical properties varying particle length and width for the same particle composition: specifically, 7.5 μm-long Fe nanorods with cross-section 0.08 μm × 0.08 μm, compared to 9 μm-long Fe nanorods with cross-section 0.16 μm × 0.16 μm. Orientation-averaged optical properties calculated using a 3D Finite-Difference Time-Domain (FDTD) approach. Upper panel: Solid black line corresponds to total extinction, dotted black line to scattering, both for 7.5 μm-long Fe nanorods with cross-section 0.08 μm × 0.08 μm. Lower panel shows scattering asymmetry. Also shown in both panels are spectra for natural dust assuming a log-gaussian particle size distribution centered on 2.5 μm (48) (red dotted lines). In both panels, gray lines correspond to the results for a 9 μm-long Fe nanorod with cross-section 0.16 μm × 0.16 μm. As expected given its spectrum, the 7.5 μm-long nanorod gives ~2× less warming per nanorod in climate simulations. However, as it uses 5× less Fe, this design is more effective on a warming-per-unit-mass basis.

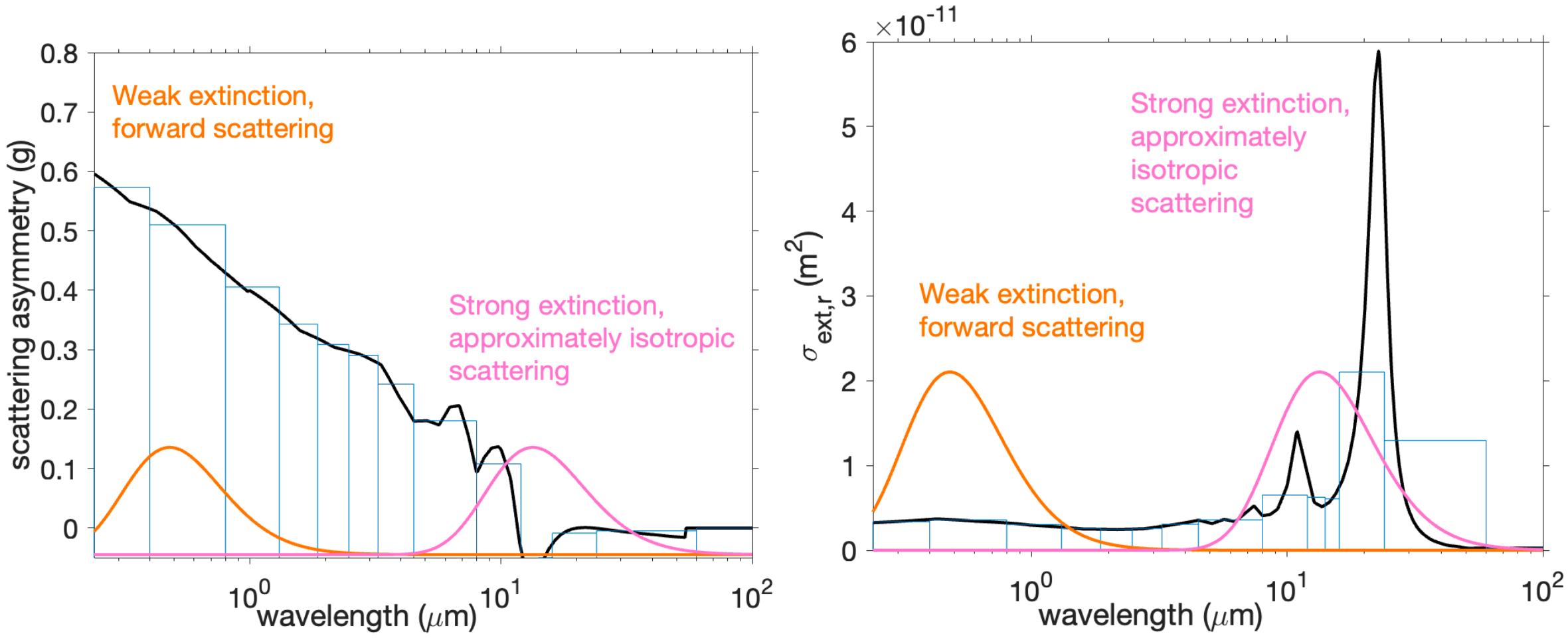

**Fig. S9.** Showing how the spectra are represented within the 3-D climate model. 9 μm-long, ~60:1 aspect ratio nanorod. 4.5 μm marks the separation between solar bins (Planck-weighted using the orange 6000 K blackbody curve, normalized W/m$^2$/μm flux density), and thermal IR bins (Planck-weighted using the pink 215 K blackbody curve, normalized W/m$^2$/μm flux density). The relative heights of the orange and pink Planck functions are to guide the eye only and have no physical significance.

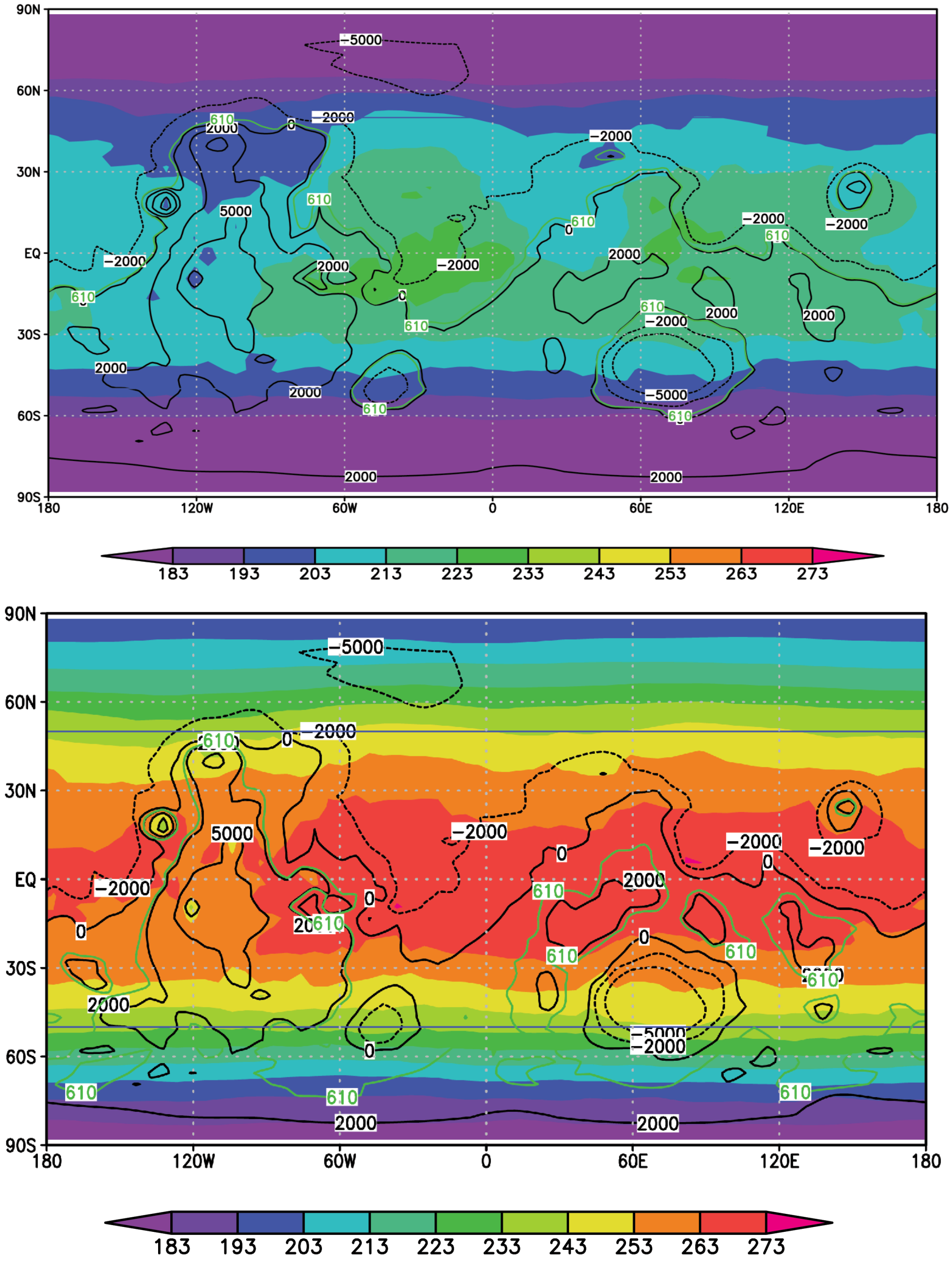

**Fig. S10.** Annual average temperatures (K) for (top) control simulation without nanorods (i.e., Fig. 2b), and (bottom) warmed case with ~160 mg/m$^2$ Al nanorods (i.e., Fig. 2a). Labeled black contours correspond to topographic elevation in m. Green contour corresponds to 610 Pa (~6 mbar) mean pressure level. Blue lines: approximate latitudinal (equatorward) extent of ice at <1 m depths.

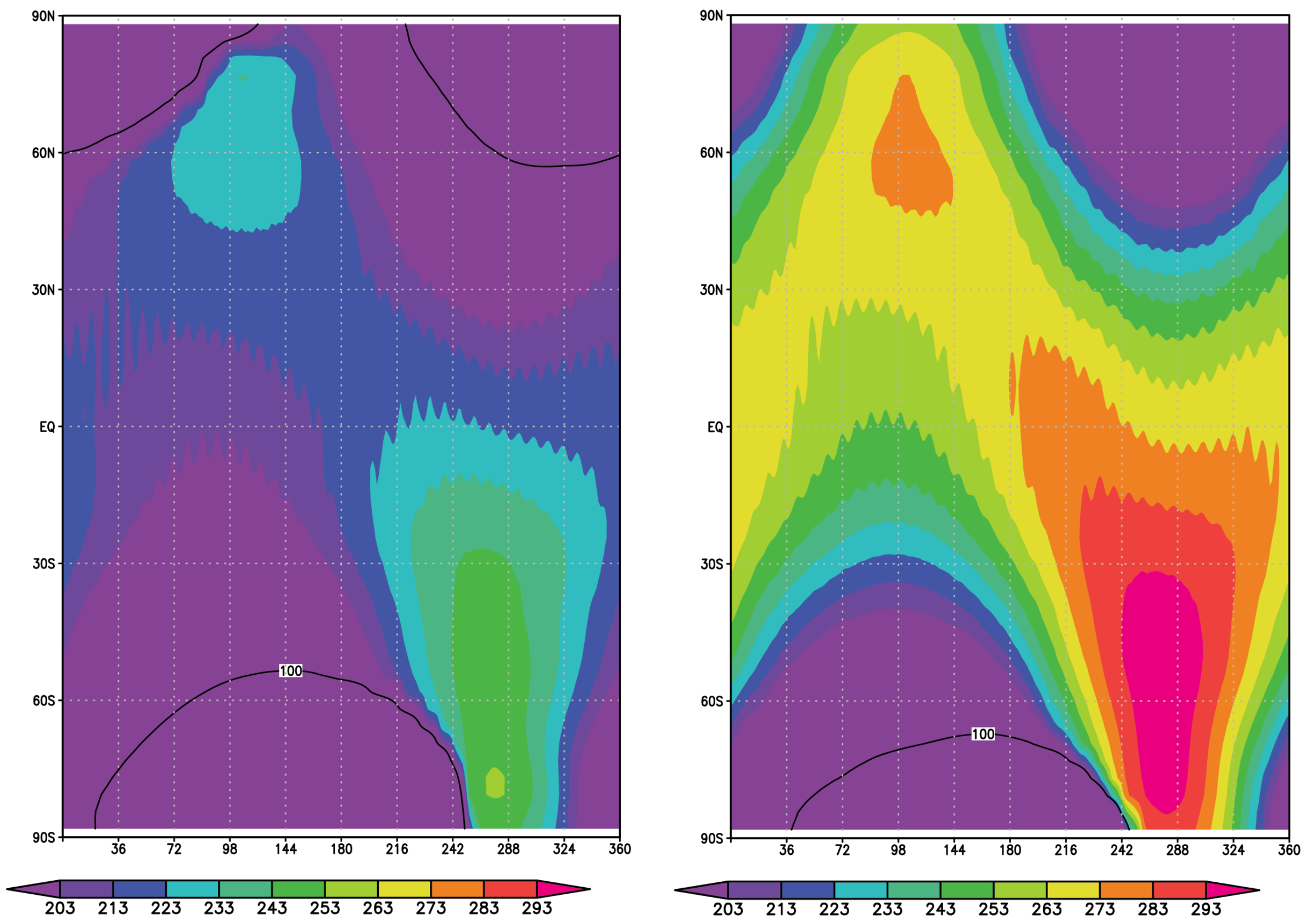

**Fig. S11.** Seasonal variation of diurnal-average surface temperature (K) for (left) control simulation without nanorods (i.e., Fig. 2b), and (right) warmed case with ~160 mg/m$^2$ nanorods (i.e., Fig. 2a), showing longitudinal average surface temperature. Each of the 10 increments on the x-axis, which are equally spaced in time, corresponds to 1/10 of a Mars year (69 Earth days). To make this figure, a 9-point smoother has been used (3 points in time, and 3 points in latitude) in order to damp oscillations associated with aliasing in the sampling of the day-night temperature cycle in the output. The black lines show the limit of substantial polar seasonal $CO_2$ ice.

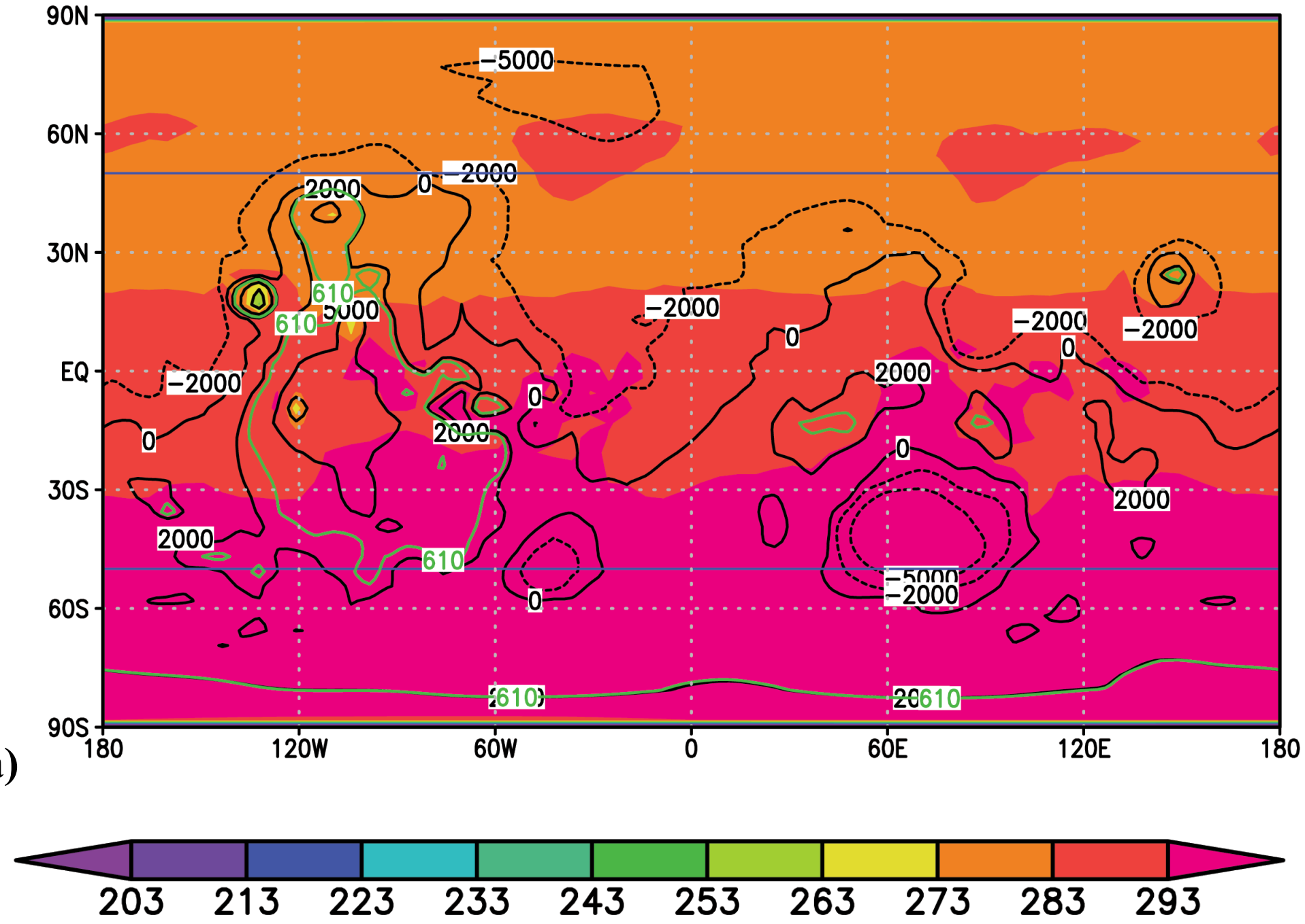

(a)

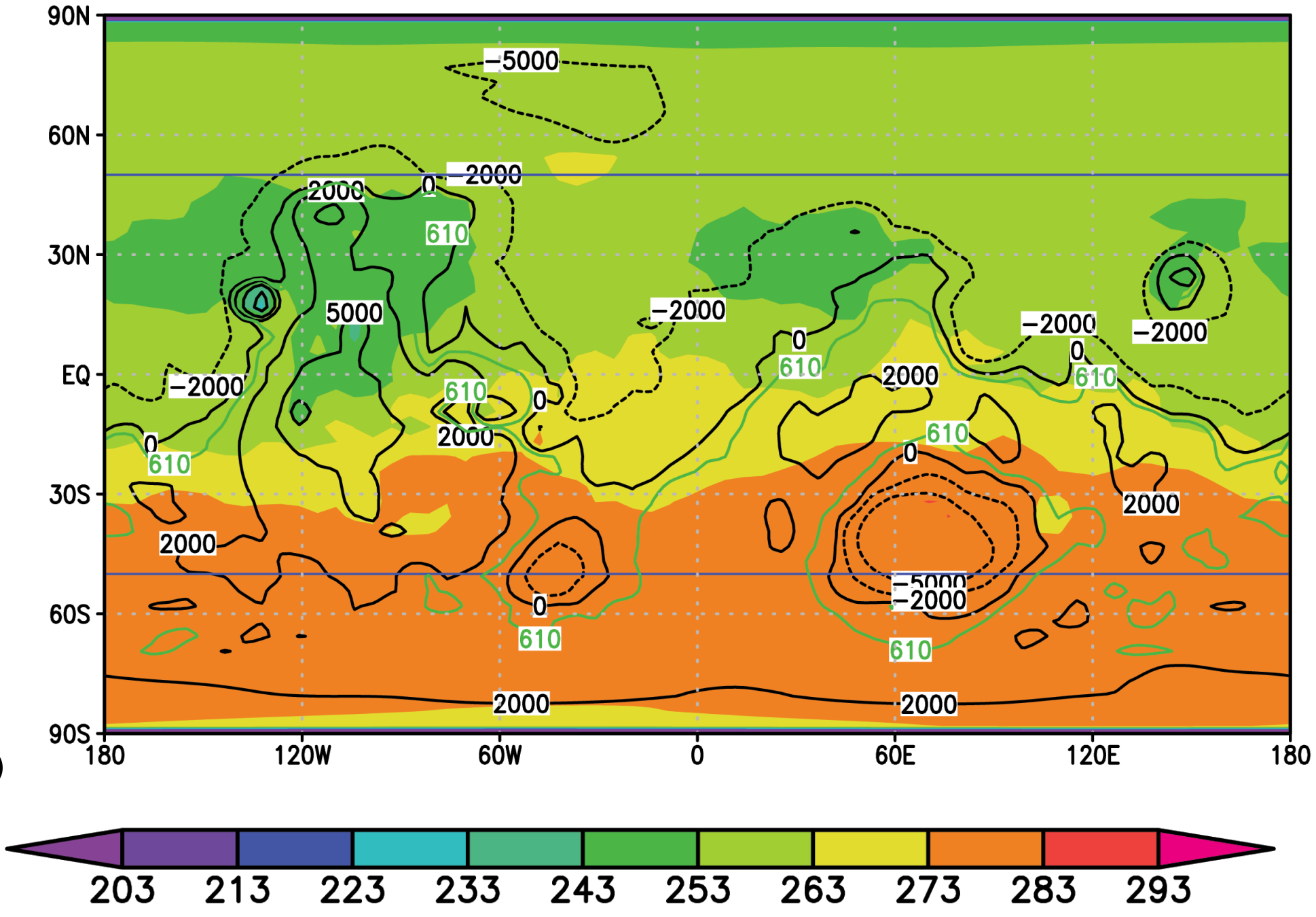

(b)

**Fig. S12.** Sensitivity tests showing how warm-season temperatures vary for different parameter choices. Baseline is Fe nanorod results shown in Fig. S19. Labeled black contours correspond to topographic elevation in m. Green contour corresponds to 610 Pa (~6 mbar) mean pressure level. Blue lines: approximate latitudinal (equatorward) extent of ice at <1 m depths. (a) More nanorods ($\tau_{Fe}$ = 1.5): the average surface pressure exceeds 610 Pa (green contour) at all locations below +2 km elevation. (b) Fewer nanorods ($\tau_{Fe}$ = 0.375).

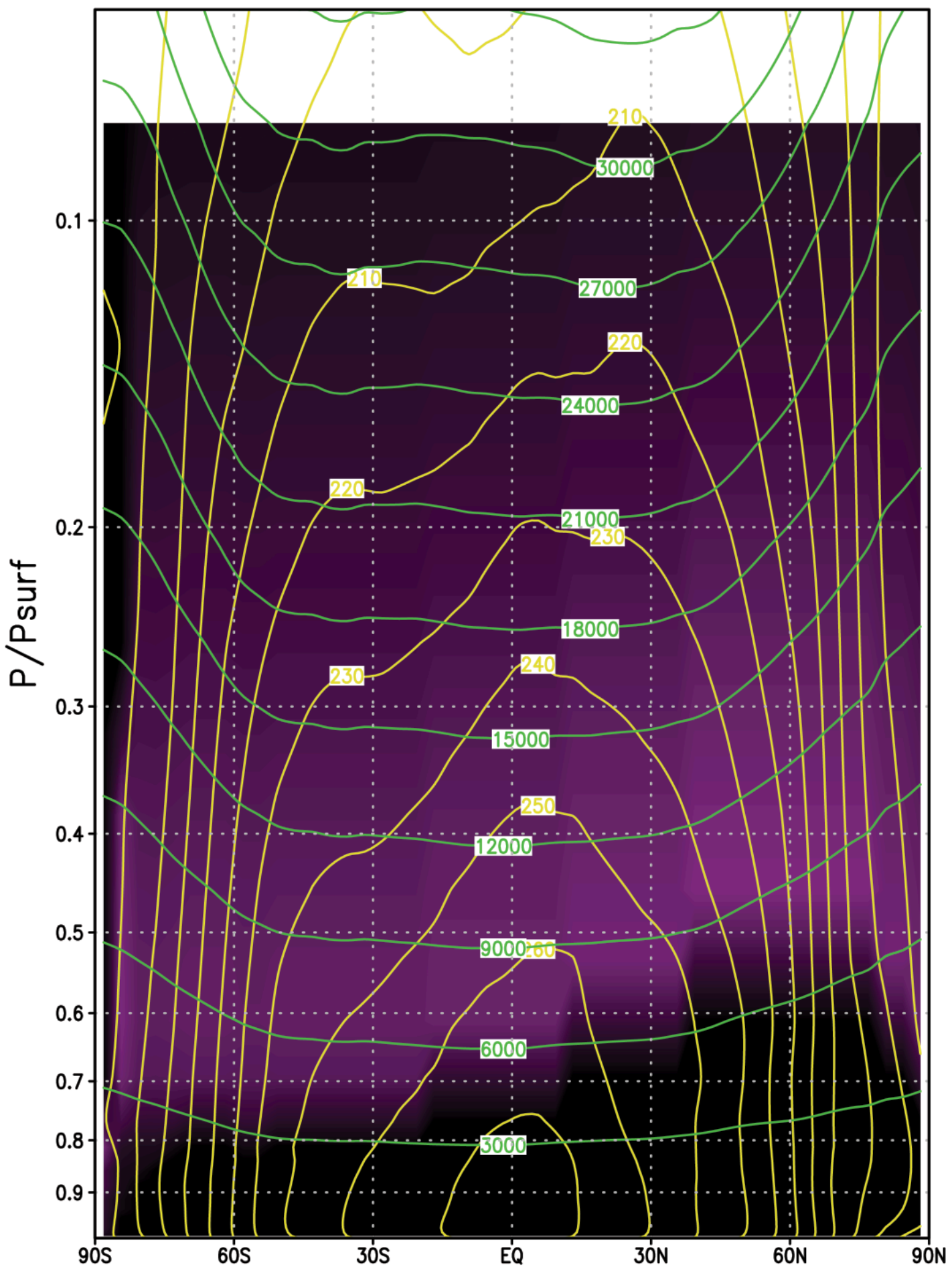

**Fig. S13.** Cross section of steady (imposed) nanorod layer ($\tau_{Fe}$ = 0.75, Fig. S19). The purple shading corresponds to the relative volume density of nanorods. The mixing ratio of nanorods within the purple layer is uniform, but there is more gas closer to the planet's surface so there are also more nanorods lower down. The tenuous part of the nanorod layer extending to high altitude is disproportionately important because of its strong contribution to the greenhouse effect. The Y-axis uses terrain-following η-coordinates ($\eta = P/P_{surf}$, where the model-top pressure has been subtracted from both pressures), and the cloud lines are tilted down and to the left because the southern hemisphere surface is topographically higher and thus closer to the fixed-pressure base of the nanorod layer. The yellow lines are contours of annual average atmospheric temperature. The green lines are contours of altitude in meters, which bow upwards at the poles where low temperatures compact the gas column.

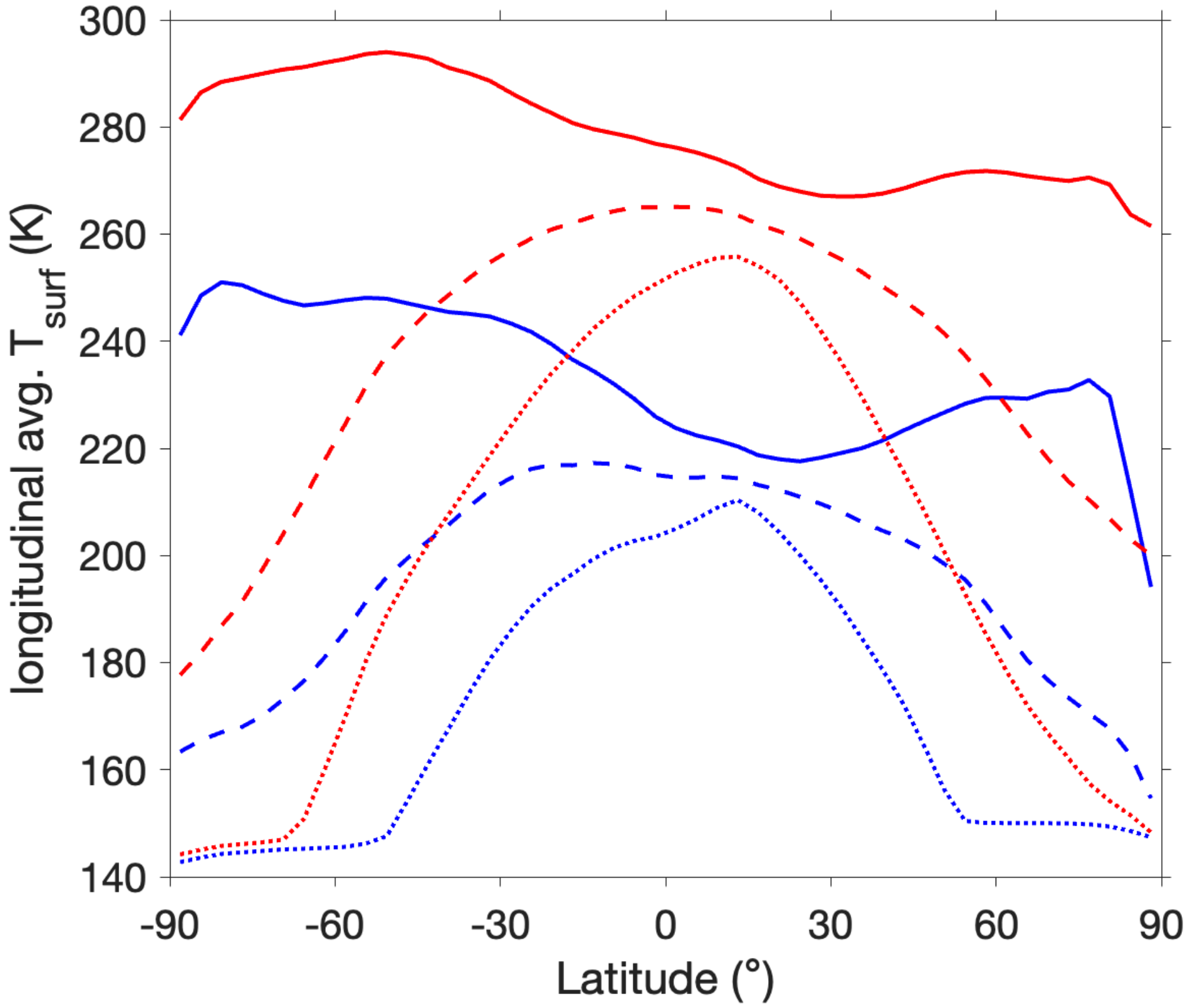

**Fig. S14.** Summary of the latitudinal and seasonal dependence of surface temperature for (red lines) the 3D model with-nanorods $\tau_{Fe}$ = 0.75 case (Fig. S19a), and (blue lines) the 3D model without nanorods (Fig. S19b). Solid line corresponds to the average temperature during the warmest season (~70 day period) during the year, dotted line corresponds to the average temperature during the coldest season (~70 day period) of the year, and dashed line corresponds to the annual average. The flattening-out of the lines around 145 K corresponds to buffering at the frost point of $CO_2$.

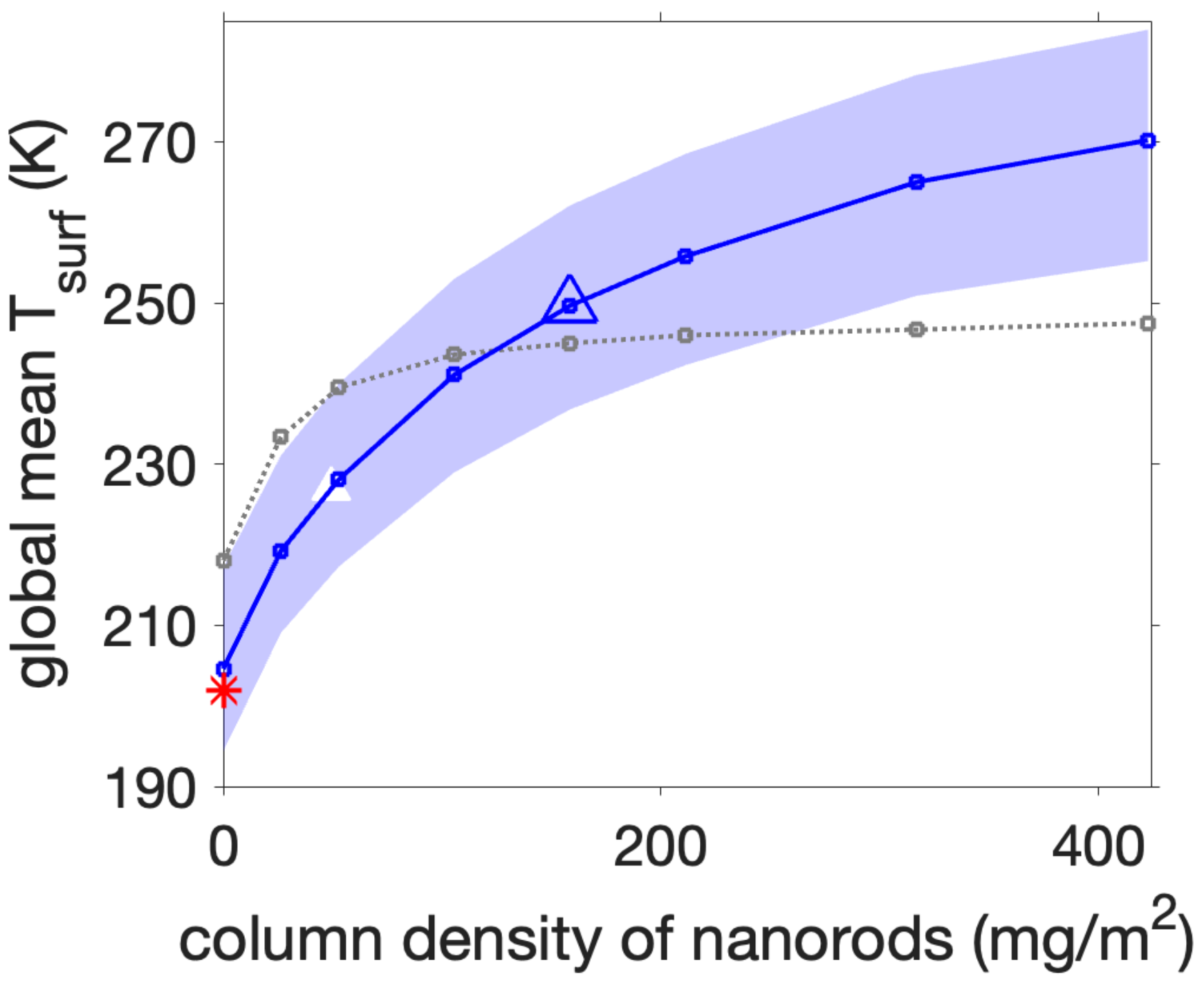

**Fig. S15.** As Fig 2c, but with a linear x-axis instead of a logarithmic x-axis. Dependence of planet-averaged surface warming on nanorod column mass. Blue triangle corresponds to Fig. 2a, the intersection of the blue line with the zero-nanorods axis corresponds to Fig. 2b, and white triangle marks onset of warm-season temperatures above the freezing point of water at 50°S. Blue corresponds to 3-D results. The blue envelope corresponds to the modeled seasonal range in global mean $T_{surf}$. Gray corresponds to 1-D results. The red asterisk corresponds to the observed modern Mars value.

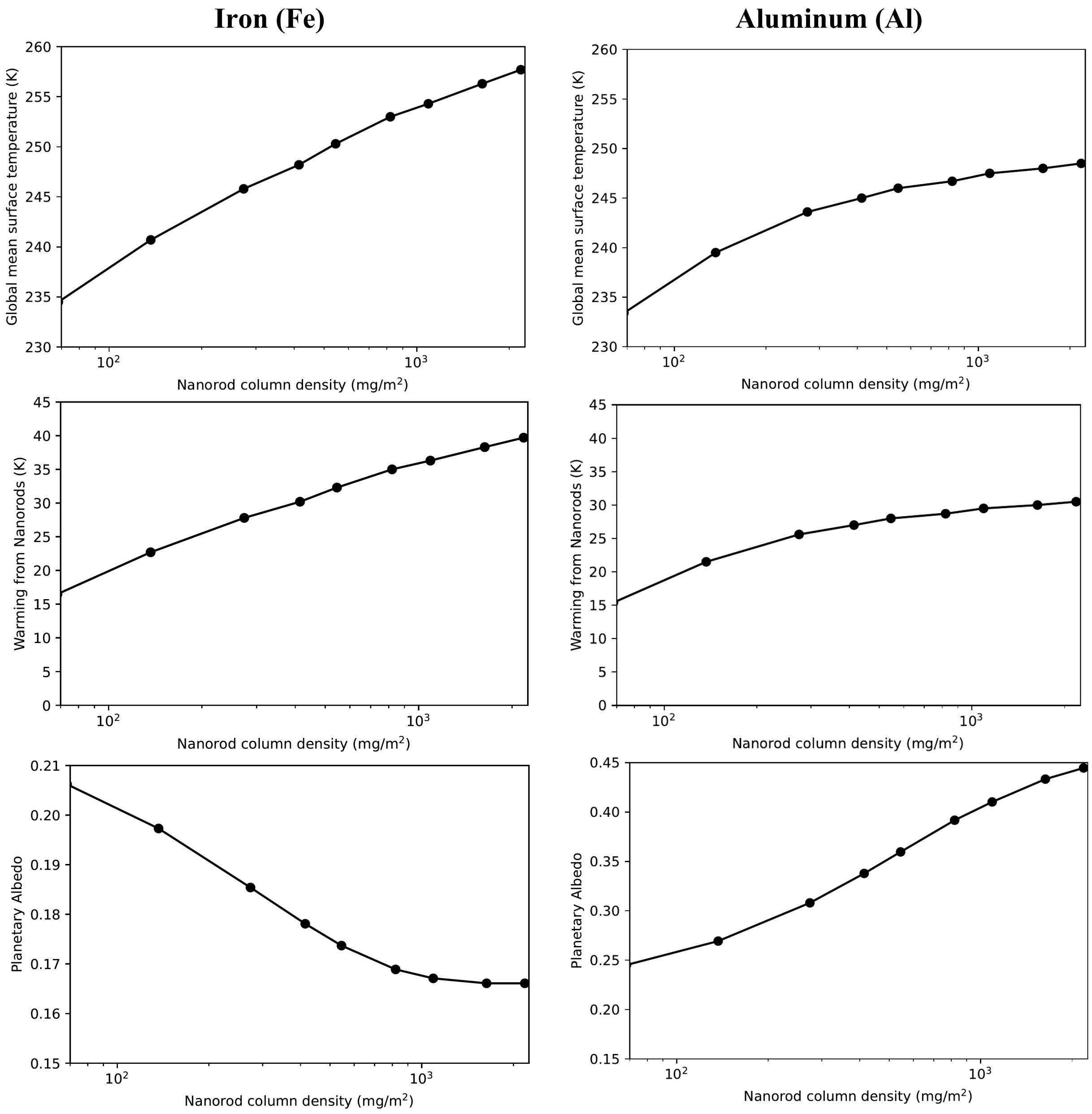

**Fig. S16.** Summary of 1-D climate model output as a function of nanorod column density. Left column corresponds to Fe, right column corresponds to Al.

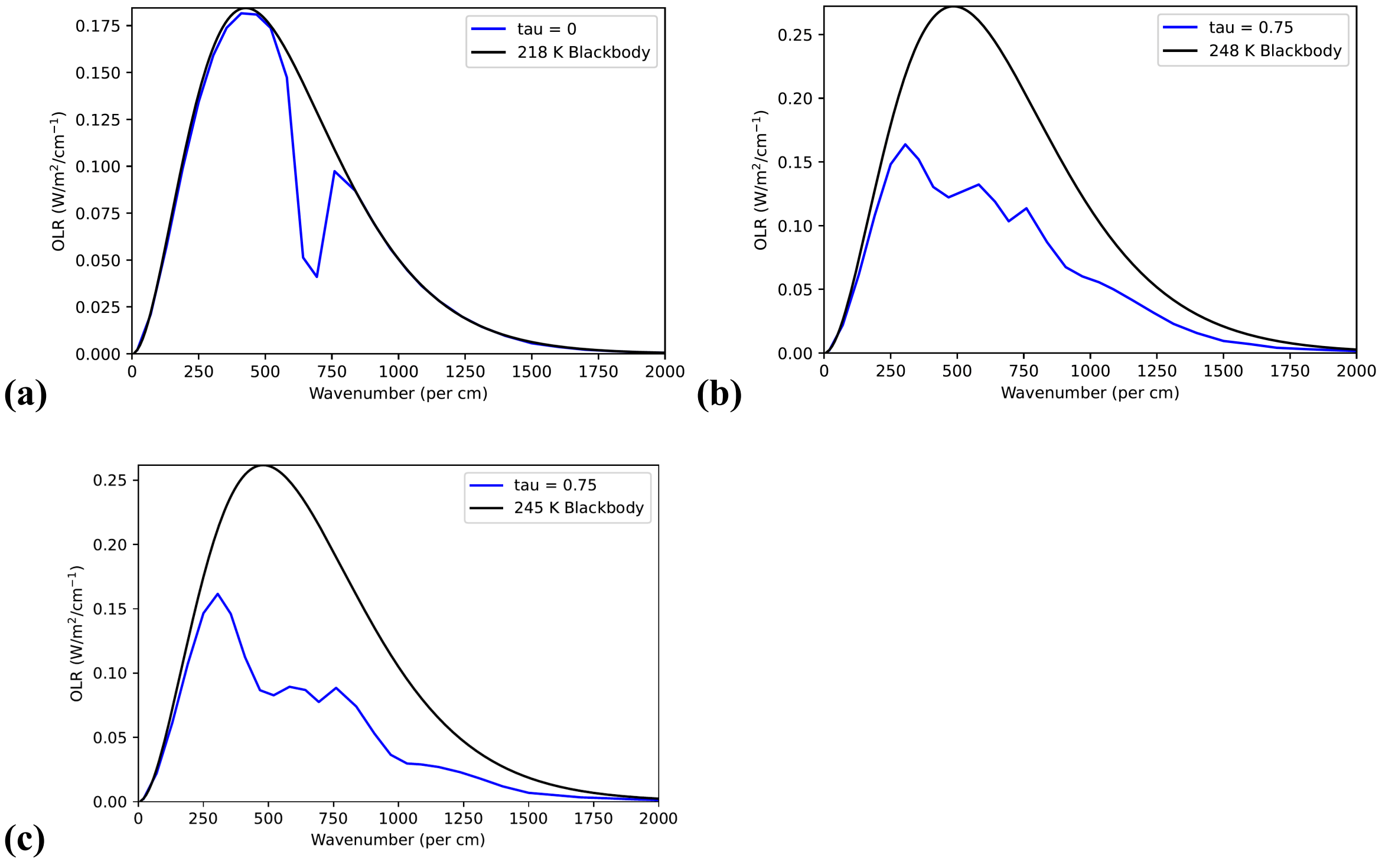

**Fig. S17.** Outgoing Longwave Radiation (OLR) from the 1D model. The integrated top-of-atmosphere OLR values are ~114.6, 96.6, and 120 W/m$^2$ for the no-nanorod, Al and Fe nanorod cases, respectively. **(a)** The no-nanorods case. Spectral windows are visible as peaks on either side of the $CO_2$ band which is at 600-800 per cm. **(b)** The nanorods case with $\tau_{Fe} = 0.75$. Absence of prominent peaks corresponds to closure of the spectral windows. **(c)** The nanorods case with $\tau_{Al} = 0.75$.

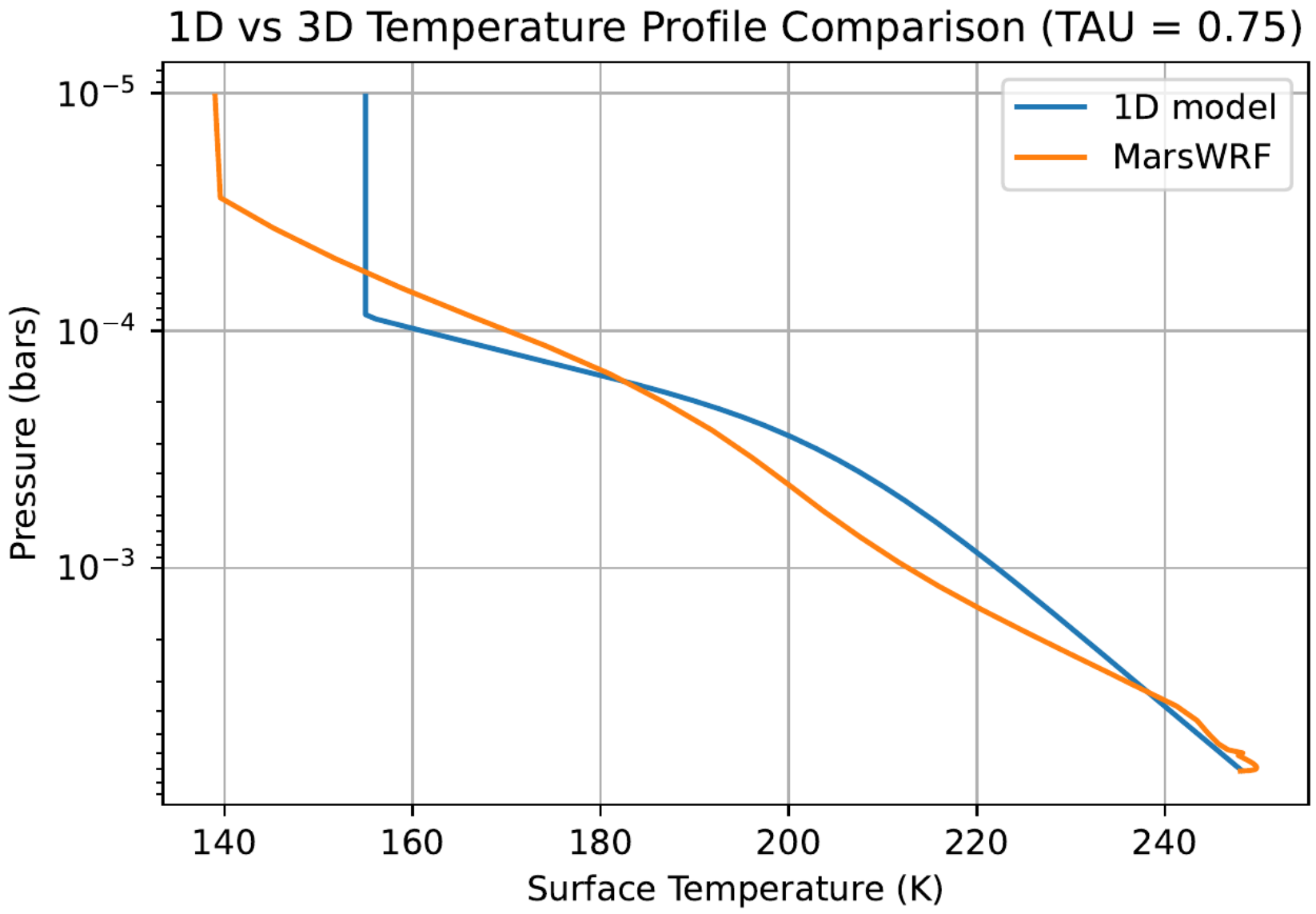

**Fig. S18.** 1D model vs. 3D (MarsWRF) model temperature profile comparison for $\tau_{Fe}$ = 0.75. The 3D result is the global and annual average. At this optical depth, surface temperature for both models is ~ 250 K.

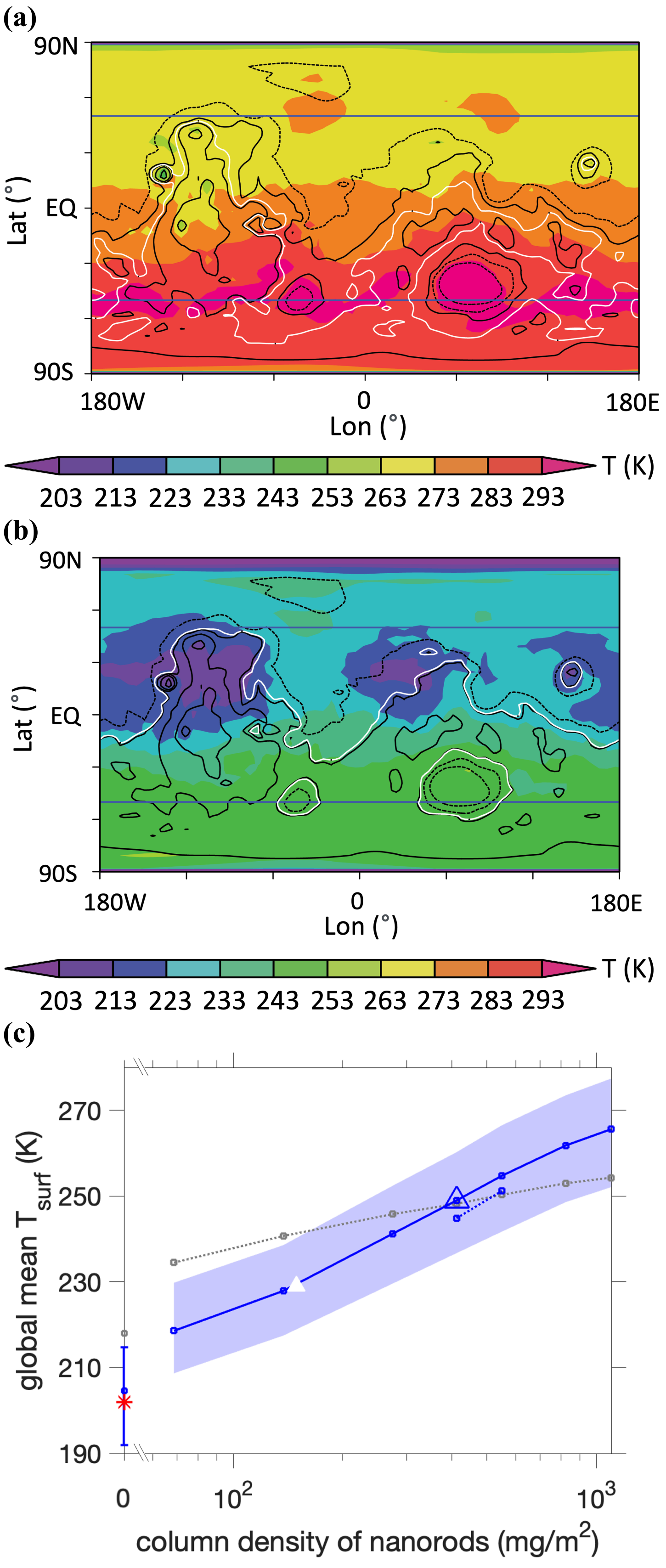

**Fig. S19.** As Fig. 2, but for 400 mg/m$^2$ loading of Fe nanorods instead of 160 mg/m$^2$ loading of Al nanorods. Warm-season temperatures (K) (color shading) on (a) Mars with addition of ~400 mg/m$^2$ of nanorods, (b) control case. This corresponds to the average surface temperature during the warmest 36° of solar longitude (~70 days) of the year. White contour corresponds to 610 Pa (~6 mbar) mean pressure level. Black contours correspond to topographic elevation in m (dashed: -5 km and –2 km, solid: 0 km, +2km, and +5 km). Blue lines: approximate latitudinal (equatorward) extent of ice at <1 m depths. Results do not include $CO_2$ outgassing from within polar ice, which would cause further warming. (c) Dependence of planet-averaged surface warming on nanorod column mass. Blue line corresponds to 3-D results, varying layer-top height between ~35 km (solid line) and ~28 km (dashed line). The blue envelope corresponds to the modeled seasonal range in global mean $T_{surf}$. Gray corresponds to 1-D results (see text for details). Blue triangle corresponds to panel (a), and white triangle marks onset of warm-season temperatures above the freezing point of water at 50°S. Symbols on y-axis are temperatures for the no-nanorod case, with the red asterisk corresponding to observed Mars value. More detail is shown in Figs. S10-S17.

| Material | Iron (Fe) | | | Aluminum (Al) | | |
|---|---|---|---|---|---|---|
| wavelength ($\times 10^{-4}$ m) | absorption cross-section ($\times 10^{-10}$ $m^2$) | scattering cross-section ($\times 10^{-10}$ $m^2$) | asymmetry parameter (g) | absorption cross-section ($\times 10^{-10}$ $m^2$) | scattering cross-section ($\times 10^{-10}$ $m^2$) | asymmetry parameter (g) |
| 0.0024 | 0.0131 | 0.0194 | 0.596 | 0.0114 | 0.0236 | 0.5204 |
| 0.0033 | 0.0133 | 0.0215 | 0.549 | 0.0053 | 0.0325 | 0.4205 |
| 0.0043 | 0.0138 | 0.0232 | 0.5327 | 0.0044 | 0.0351 | 0.4454 |
| 0.0052 | 0.0131 | 0.0226 | 0.505 | 0.0036 | 0.0298 | 0.4093 |
| 0.0061 | 0.0128 | 0.0214 | 0.4728 | 0.0034 | 0.0274 | 0.3687 |
| 0.007 | 0.0128 | 0.0205 | 0.4492 | 0.0037 | 0.0254 | 0.3548 |
| 0.008 | 0.0127 | 0.0196 | 0.4301 | 0.0045 | 0.0233 | 0.3536 |
| 0.0089 | 0.0125 | 0.0186 | 0.4131 | 0.0044 | 0.0220 | 0.3450 |
| 0.0098 | 0.0121 | 0.0177 | 0.3979 | 0.0028 | 0.0216 | 0.3350 |
| 0.01 | 0.0125 | 0.017 | 0.3997 | 0.0012 | 0.0221 | 0.3178 |
| 0.0158 | 0.0104 | 0.0148 | 0.3322 | 0.0011 | 0.0213 | 0.2824 |
| 0.0217 | 0.0098 | 0.0147 | 0.3041 | 0.0022 | 0.0274 | 0.2659 |
| 0.0275 | 0.0102 | 0.0152 | 0.2921 | 0.0032 | 0.0349 | 0.2400 |
| 0.0333 | 0.012 | 0.0168 | 0.2745 | 0.0048 | 0.0441 | 0.0693 |
| 0.0392 | 0.0127 | 0.0193 | 0.2218 | 0.0068 | 0.0562 | 0.0755 |
| 0.045 | 0.0163 | 0.0191 | 0.1798 | 0.0027 | 0.0303 | 0.1405 |
| 0.0508 | 0.0112 | 0.0203 | 0.1804 | 0.0043 | 0.0394 | 0.1514 |
| 0.0567 | 0.0152 | 0.0213 | 0.1732 | 0.0025 | 0.0377 | 0.1627 |
| 0.0625 | 0.012 | 0.0217 | 0.2035 | 0.0054 | 0.0532 | 0.1700 |
| 0.0683 | 0.0163 | 0.0271 | 0.206 | 0.0073 | 0.0588 | 0.0671 |
| 0.0742 | 0.0214 | 0.031 | 0.1553 | 0.0044 | 0.0443 | 0.0308 |
| 0.08 | 0.013 | 0.0278 | 0.0922 | 0.0046 | 0.0508 | 0.0664 |
| 0.0858 | 0.012 | 0.0302 | 0.119 | 0.0053 | 0.0650 | 0.0825 |
| 0.0917 | 0.0138 | 0.0347 | 0.1344 | 0.0101 | 0.0916 | 0.0818 |
| 0.0975 | 0.0225 | 0.0452 | 0.1368 | 0.0182 | 0.1456 | 0.0490 |
| 0.1 | 0.0247 | 0.0514 | 0.1327 | 0.0223 | 0.1579 | 0.0321 |
| 0.1092 | 0.0532 | 0.0886 | 0.0835 | 0.0130 | 0.0854 | -0.0657 |
| 0.1184 | 0.0359 | 0.0619 | -0.0233 | 0.0066 | 0.0549 | -0.1176 |
| 0.1276 | 0.0165 | 0.0399 | -0.0872 | 0.0057 | 0.0551 | -0.1074 |
| 0.1367 | 0.0125 | 0.0387 | -0.0822 | 0.0064 | 0.0630 | -0.0808 |
| 0.1459 | 0.0126 | 0.0437 | -0.057 | 0.0080 | 0.0768 | -0.0579 |
| 0.1551 | 0.0145 | 0.0524 | -0.0352 | 0.0107 | 0.0976 | -0.0427 |
| 0.1643 | 0.0178 | 0.0644 | -0.0205 | 0.0146 | 0.1274 | -0.0337 |
| 0.1735 | 0.0234 | 0.0813 | -0.0109 | 0.0215 | 0.1720 | -0.0284 |
| 0.1827 | 0.032 | 0.1081 | -0.0053 | 0.0327 | 0.2449 | -0.0254 |
| 0.1918 | 0.0451 | 0.1489 | -0.0018 | 0.0501 | 0.3503 | -0.0240 |
| 0.201 | 0.067 | 0.2132 | -0.0001 | 0.0700 | 0.4696 | -0.0235 |
| 0.2102 | 0.0987 | 0.3056 | 0.0006 | 0.0818 | 0.5050 | -0.0235 |
| 0.2194 | 0.1371 | 0.4119 | 0.0007 | 0.0707 | 0.4157 | -0.0240 |
| 0.2286 | 0.1553 | 0.4378 | 0.0003 | 0.0501 | 0.2839 | -0.0246 |
| 0.2378 | 0.1293 | 0.3514 | -0.0004 | 0.0349 | 0.1829 | -0.0252 |
| 0.2469 | 0.0886 | 0.2335 | -0.0011 | 0.0246 | 0.1214 | -0.0259 |
| 0.2561 | 0.0604 | 0.1484 | -0.0019 | 0.0169 | 0.0840 | -0.0268 |
| 0.2653 | 0.0428 | 0.099 | -0.0028 | 0.0124 | 0.0599 | -0.0276 |
| 0.2745 | 0.0304 | 0.0691 | -0.0037 | 0.0101 | 0.0443 | -0.0283 |
| 0.2837 | 0.0228 | 0.0495 | -0.0044 | 0.0082 | 0.0340 | -0.0289 |
| 0.2929 | 0.0186 | 0.0369 | -0.0051 | 0.0065 | 0.0267 | -0.0296 |
| 0.302 | 0.0154 | 0.0285 | -0.0059 | 0.0053 | 0.0213 | -0.0303 |
| 0.3112 | 0.0124 | 0.0224 | -0.0066 | 0.0047 | 0.0172 | -0.0309 |
| 0.3204 | 0.0103 | 0.0179 | -0.0072 | 0.0042 | 0.0141 | -0.0314 |
| 0.3296 | 0.0089 | 0.0145 | -0.0077 | 0.0038 | 0.0117 | -0.0318 |
| 0.3388 | 0.008 | 0.0119 | -0.0081 | 0.0033 | 0.0098 | -0.0321 |
| 0.348 | 0.007 | 0.0099 | -0.0085 | 0.0028 | 0.0083 | -0.0325 |
| 0.3571 | 0.0061 | 0.0084 | -0.009 | 0.0026 | 0.0071 | -0.0329 |
| 0.3663 | 0.0054 | 0.0071 | -0.0095 | 0.0024 | 0.0061 | -0.0332 |
| 0.3755 | 0.0049 | 0.0061 | -0.01 | 0.0022 | 0.0053 | -0.0335 |
| 0.3847 | 0.0045 | 0.0052 | -0.0104 | 0.0020 | 0.0046 | -0.0338 |
| 0.3939 | 0.0041 | 0.0046 | -0.0108 | 0.0017 | 0.0040 | -0.0340 |
| 0.4031 | 0.0036 | 0.004 | -0.0112 | 0.0015 | 0.0036 | -0.0342 |
| 0.4122 | 0.0032 | 0.0035 | -0.0115 | 0.0014 | 0.0032 | -0.0343 |
| 0.4214 | 0.0029 | 0.0031 | -0.0119 | 0.0013 | 0.0028 | -0.0345 |
| 0.4306 | 0.0027 | 0.0028 | -0.0123 | 0.0014 | 0.0025 | -0.0348 |
| 0.4398 | 0.0027 | 0.0025 | -0.0127 | 0.0014 | 0.0023 | -0.0350 |
| 0.449 | 0.0027 | 0.0022 | -0.0132 | 0.0014 | 0.0021 | -0.0352 |
| 0.4582 | 0.0026 | 0.002 | -0.0136 | 0.0013 | 0.0019 | -0.0353 |
| 0.4673 | 0.0024 | 0.0018 | -0.0139 | 0.0011 | 0.0017 | -0.0355 |
| 0.4765 | 0.0021 | 0.0017 | -0.0143 | 0.0009 | 0.0016 | -0.0356 |
| 0.4857 | 0.0018 | 0.0015 | -0.0146 | 0.0007 | 0.0014 | -0.0358 |
| 0.4949 | 0.0015 | 0.0014 | -0.0149 | 0.0005 | 0.0013 | -0.0359 |
| 0.5041 | 0.0013 | 0.0012 | -0.0151 | 0.0003 | 0.0012 | -0.0359 |
| 0.5133 | 0.0012 | 0.0011 | -0.0153 | 0.0003 | 0.0011 | -0.0360 |
| 0.5224 | 0.0012 | 0.001 | -0.0155 | 0.0004 | 0.0010 | -0.0361 |
| 0.5316 | 0.0013 | 0.0009 | -0.0156 | 0.0006 | 0.0009 | -0.0361 |
| 0.5408 | 0.0015 | 0.0009 | -0.0158 | 0.0008 | 0.0008 | -0.0362 |
| 0.55 | 0.0017 | 0.0008 | -0.016 | 0.0010 | 0.0008 | -0.0363 |

**Supplementary Table 1:** Orientation-averaged optical properties of nanorods as used in the climate simulations, corresponding to a 9 μm-long nanorod with cross-section (0.16×0.16) μm.

| Optical depth ($\tau$) at $\lambda$ = 0.67 μm | Nanorod column mass (mg/m$^2$) | 1-D temperature output, global mean (K) | 3D temperature output | | |
|---|---|---|---|---|---|
| | | | 3-D global mean temperature (K) | 50°N warm-season temperature (K) | 50°S warm-season temperature (K) |
| 0 | 0 | 218.0 | 204.6 | 228.8 | 250.3 |
| 0.125 (Al) | 26.6 | 233.4 | 219.2 | 243.2 | 264.3 |
| 0.25 (Al) | 53.2 | 239.5 | 228.1 | 253.0 | 274.4 |
| 0.5 (Al) | 106.5 | 243.6 | 241.1 | 265.8 | 289.0 |
| 0.75 (Al) | 161.2 | 245.0 | 249.7 | 274.3 | 298.2 |
| 1 (Al) | 212.1 | 246.0 | 255.8 | 280.2 | 304.8 |
| 1.5 (Al) | 319.4 | 246.7 | 265.0 | 287.6 | 312.3 |
| 2 (Al) | 424.6 | 247.5 | 270.3 | 291.6 | 316.7 |
| 0.125 (Fe) | 68.4 | 234.5 | 218.6 | 241.2 | 262.2 |
| 0.25 (Fe) | 136.6 | 240.7 | 227.9 | 250.9 | 271.9 |
| 0.5 (Fe) | 273.8 | 245.8 | 241.2 | 263.2 | 285.9 |
| 0.75 (Fe) | 414.3 | 248.2 | 249.0 | 270.9 | 294.0 |
| 1 (Fe) | 545.1 | 250.3 | 254.7 | 275.7 | 299.3 |
| 1.5 (Fe) | 820.8 | 253.0 | 261.8 | 281.5 | 304.9 |
| 2 (Fe) | 1091.1 | 254.3 | 265.6 | 285.0 | 307.1 |

**Supplementary Table 2.** Summary of climate model output. The nanorod column mass differs between the Al and Fe cases by a factor of 2.57 (ratio of material densities, corrected for the 13% greater extinction cross-section of Fe rods at 0.67 μm relative to Al rods).

| Description | Optical depth ($\tau$) at $\lambda$ = 0.67 µm | 3D temperature output | | |
|---|---|---|---|---|
| | | 3-D global mean temperature (K) | 50°N warm-season temperature (K) | 50°S warm-season temperature (K) |
| *Reference* | *0.75* | *249.0* | *270.9* | *294.0* |
| Lower cloud top (~28 km vs. ~35 km) | 0.75 | 244.8 | 270.6 | 289.5 |
| Lower cloud top (~28 km vs. ~35 km) | 1 | 251.2 | 275.6 | 297.1 |
| 2× atmospheric pressure (≈12 mbar) | 0.75 | 252.7 | 269.9 | 290.1 |
| Nanorods extend only from 45°S - 45°N | 0.75 | 237.7 | 230.5 | 253.2 |
| Nanorods only active from $L_s$ = 180 – 360, passive at other seasons | 0.75 | 224.9 | 229.2 | 293.3 |
| 5× greater output frequency | 0.75 | 249.0 | 270.8 | 294.0 |
| 2× smaller numerical timestep | 0.75 | 249.5 | 271.0 | 293.5 |
| Thinner cloud (200 Pa base, 8 levels) | 0.75 | 248.7 | 269.3 | 290.9 |
| 96×72 spatial grid in model (vs. default of 64×48) | 0.75 | 250.5 | 271.1 | 292.9 |

**Supplementary Table 3.** Summary output for additional sensitivity tests using 3D model. All were carried out using Fe nanorods.